\documentclass[twocolumn]{aastex62}

\usepackage{amssymb}
\usepackage{color}
\usepackage{amsmath}
\usepackage{comment}
\usepackage{lipsum}
\usepackage{hyperref}
\usepackage{soul}
\usepackage{graphicx}
\usepackage{natbib}
\usepackage{mathtools}
\usepackage{lipsum}
\usepackage{xcolor}
\usepackage{bm}

\DeclarePairedDelimiter\abs{\lvert}{\rvert}
\newcommand{\lens}{\bm{L}}
\newcommand{\src}{\bm{s}}
\newcommand{\data}{\bm{d}}
\newcommand{\noise}{\bm{n}}
\newcommand{\cov}{\bm{C}_\textrm{d}}

\newcommand{\reg}{\bm{R}}

\newcommand{\HtwoD}{\bm{H_{\text{2D}}}}
\newcommand{\Lhat}{\bm{\hat{L}}}
\newcommand{\Chat}{\bm{\hat{C}}_\textrm{d}}

\newcommand{\RtwoDhat}{\bm{\hat{R}_{\text{2D}}}}
\newcommand{\RthreeDhat}{\bm{\hat{R}_{\text{3D}}}}

\newcommand{\HtwoDhat}{\bm{\hat{H}_{\text{2D}}}}
\newcommand{\HthreeDhat}{\bm{\hat{H}_{\text{3D}}}}
\newcommand{\RtwoD}{\bm{R_{\text{2D}}}}
\newcommand{\Qhat}{\bm{\hat{Q}}}
\newcommand{\Ahat}{\bm{\hat{A}}}
\newcommand{\Fhat}{\bm{\hat{F}}}
\newcommand{\blur}{\bm{B}}
\newcommand{\blurhat}{\bm{\hat{B}}}
\newcommand{\Fmat}{\bm{F}}
\newcommand{\Amat}{\bm{A}}
\newcommand{\Dmat}{\bm{D}}
\newcommand{\Mmat}{\bm{M}}
\newcommand{\Identity}{\bm{I}}

\newcommand{\kms}{\text{km}\,\text{s}^{-1}}
\newcommand{\code}[1]{\texttt{#1}}

\graphicspath{{./}{figures/}}

\received{November 14, 2021}
\revised{February 24, 2022}
\accepted{February 28, 2022}
\published{April 8, 2022}

\submitjournal{ApJ}

\shorttitle{Reconstruction of Strongly Lensed Galaxies with Resolved Kinematics}
\shortauthors{Young et al.}

\usepackage{layouts}

\begin{document}
	\title{A New Method for the Reconstruction of Strongly Lensed Galaxies with Resolved Kinematics}
	\correspondingauthor{A.~J. Young}
	\email{ayoung@physics.rutgers.edu}
	
	\author[0000-0002-3053-501X]{A.~J. Young}
	\affiliation{Department of Physics and Astronomy, Rutgers, the State University of New Jersey, 136 Frelinghuysen Road, Piscataway, NJ 08854-8019, USA}
	
	\author[0000-0001-6812-2467]{C.~R. Keeton}
	\affiliation{Department of Physics and Astronomy, Rutgers, the State University of New Jersey, 136 Frelinghuysen Road, Piscataway, NJ 08854-8019, USA}
	
	\author[0000-0002-7892-396X]{A.~J. Baker}
	\affiliation{Department of Physics and Astronomy, Rutgers, the State University of New Jersey, 136 Frelinghuysen Road, Piscataway, NJ 08854-8019, USA}
	
	\begin{abstract}
		Integral field spectroscopy of high-redshift galaxies has become a powerful tool for understanding their dynamics and evolutionary states. However, in the case of gravitationally lensed systems, it has proved difficult to model both lensing and intrinsic kinematics in a way that takes full advantage of the information available in the spectral domain. In this paper, we introduce a new method for pixel-based source reconstruction that alters standard regularization schemes for two-dimensional data in a way that leverages kinematic information in a physically motivated but flexible fashion, and that is better suited to the three-dimensional nature of integral field data. To evaluate the performance of this method, we compare its results to those of a more traditional two-dimensional non-parametric approach using mock ALMA observations of a typical high-redshift dusty star-forming galaxy. We find that 3D regularization applied to an entire data cube reconstructs a source's intensity and velocity structure more accurately than 2D regularization applied to separate velocity channels. Cubes reconstructed with 3D regularization also have more uniform noise and resolution properties and are less sensitive to the signal-to-noise ratio of individual velocity channels than the results of 2D regularization. Our new approach to modeling integral field observations of lensed systems can be implemented without making restrictive a priori assumptions about intrinsic kinematics, and opens the door to new observing strategies that prioritize spectral resolution over spatial resolution (e.g., for multi-configuration arrays like ALMA).
	\end{abstract}
	

	\section{Introduction}
	Strong gravitational lensing stretches and amplifies the images of distant
	sources, making them easier to detect and resolve, and allows us to probe 
	the mass distributions of the foreground structures that lens them.  Fully 
	exploiting the scientific potential of strong lensing requires 
	the simultaneous reconstruction of {\it both} the intrinsic source-plane
	structure and the lensing potential for a given system --- a complex 
	inference problem whose solution is the main goal of gravitational lens
	modeling.  For a multiply imaged system, different features observed on the
	sky originate from the same part of the source; a good lens model will 
	recover this consistency, while a bad one will not.  As a result, the 
	most robust lens models tend to be derived from data that maximize the number
	of distinct ``features'' that can be identified, i.e., the number of 
	resolution elements spanning the observed source.  In the context of 
	imaging, this criterion translates to high angular resolution, explaining 
	why the {\it Hubble Space Telescope}, ground-based telescopes equipped with 
	adaptive optics, and very long baseline interferometric arrays have made
	important contributions to strong lensing studies.  However, it is also
	possible to satisfy this criterion using data with high {\it spectral}
	resolution, which is essentially equivalent to high angular resolution for
	sources with large velocity gradients.  As noted in previous studies of {\it
	unlensed} sources \citep[e.g.,][]{Scoville1997}, in fact, the angular 
	shifts of emission centroids between adjacent velocity channels scale as 
	$\Delta\theta \propto \Delta v\,(dv/dr)^{-1}$ --- so datasets with high 
	spectral resolution (small $\Delta v$) can provide information on scales
	smaller than the nominal angular resolution element.  Leveraging such 
	information may be especially effective at radio wavelengths, where 
	heterodyne detection delivers high spectral resolution as a matter of 
	course.

	The potential value of high spectral resolution for lens modeling only 
	adds to the scientific appeal of integral field (i.e., three-dimensional, 3D)
	observations of high-redshift galaxies.  Spatially resolved studies of
	ionized and molecular gas kinematics have been used to extract rotation 
	curves for galaxies at $z \geq 2$, which can be used to constrain their dark 
	matter density profiles and their formation histories 
	\citep[e.g.,][]{Genzel2017,Genzel2020,Rizzo2021}. Spatially 
	resolved spectral line mapping in $z \geq2$ systems can reveal the morphology
	of star formation \citep[e.g.,][]{Jones2010}, the impact of gas inflow on
	chemical enrichment \citep[e.g.,][]{Cresci2010}, and the influence of active
	galactic nuclei on their surroundings \citep[e.g.,][]{Sharon2019}.  
	When such analyses of ``data cubes'' (with one spectral and two spatial 
	dimensions) are undertaken for lensed systems, they immediately benefit from
	the higher signal-to-noise ratios (S/Ns) and higher effective linear 
	resolutions that lensing affords. However, if the rich information contained
	in the cubes can also be harnessed to improve the quality of the lens models
	themselves, the payoff will be even bigger. 
	
	This paper is motivated by the need for a new method of source-plane 
	reconstruction of intrinsically 3D datasets that more fully 
	captures the information available in those datasets in a way that previous 
	methods do not. The work builds on a long history of methods for simultaneously 
	reconstructing the source-plane structure and lens potential. 
	\citet{Kochanek1992} combined lens modeling with the Clean algorithm for 
	source reconstruction. \citet{Wallington1996} introduced the idea of using 
	regularization (a prior in the Bayesian interpretation) that effectively 
	tunes the number of source parameters and penalizes source models that are 
	too complex, specifically employing the maximum entropy method. 
	\citet{Warren2003} pointed out that making the penalty function quadratic 
	in the source surface brightness led to a fast ``semi-linear'' method. 
	Several choices of quadratic regularization schemes exist, with the most 
	common penalizing the sum of either the squared pixel values, their 
	squared gradient, or their squared Laplacian. There is no objective 
	standard for the choice of regularization scheme in general. 
	\citet{Suyu2006} compared Bayesian evidences and showed that the optimal 
	choice of prior depends on both the data and the nature of the unlensed 
	source. Non-parametric source reconstruction methods have been further 
	developed under a unified Bayesian framework \citep{Brewer2006} to include 
	adaptive and irregular source plane grids \citep{Dye2005,Vegetti2009} or 
	to reconstruct a source in a shapelet basis \citep{Birrer2015,Tagore2016}.
	
	To date, the majority of strong lensing studies have focused on 2D surface 
	brightness data. For 3D data cubes, one approach is to simply reconstruct 
	a data cube's channels as a series of 
	independent sources using the 2D techniques described above 
	\citep[e.g.,][]{Stark2008}, with no prior on how the reconstructed channels 
	should relate to each other. This approach neglects the fact that the 
	channels are, in fact, not unrelated to each other for any plausible 
	kinematic state. As we discuss below, this choice often amounts to a 
	sub-optimal regularization scheme, depending on the quality of the data. 
	For sources that are rotating disks, 
	the derived source plane channel images can, if desired, be fit with an 
	analytic rotating disk model \citep[e.g.,][]{Geach2018}. However, this and 
	other analyses of independent channel reconstructions are complicated by the 
	non-uniform resolution and noise properties in the source plane.

	In recent years, several authors have developed approaches to the 
	reconstruction of lensed sources from integral field data. 
	\citet{Patricio2018} fit observations to kinematic models that have been 
	lensed into the image plane, but only fit to the velocity maps, 
	not the full data cube. \citet{Rizzo2018} present a method for the 
	non-parametric reconstruction of a full 3D data cube in which 
	regularization makes use of a parameterized rotating disk model.
	\citet{Chirivi2020} also fit a dynamical model in the source plane, but use 
	axisymmetric Jeans modeling to fit the kinematic moments of mock integral 
	field observations of stellar absorption. These approaches all address the 
	issue of non-uniform source-plane resolution and allow for (i) simultaneous 
	optimization of lens and source parameters, and (ii) exploration of 
	degeneracies between model parameters. However, while parameterized kinematic
	models can be useful in many cases, sources with disturbed kinematics, 
	complicated morphologies, and/or multiple components can deviate 
	significantly from such simple descriptions, and require less strict 
	assumptions (e.g., merely the existence of velocity coherence) for the 
	sources' true properties to be recovered.

    Convolutional neural networks and related deep learning 
    techniques have recently begun to show promise for solving 
    these and other related problems in lens modeling. Although 
    generally only trained on data using singular isothermal 
    ellipsoid (SIE) mass distributions, models have been developed 
    to predict lens parameters and their errors from an image 
    \citep{Hezaveh2017,PerreaultLevasseur2017}. 
    Other models have also been effective at reconstructing source-plane 
    structure from lensed data, showing some degree of 
    generalizability, and even achieving better results than 
    maximum likelihood methods \citep{Morningstar2019}. 
    Because the choice of training data for these networks imposes 
    a prior on the reconstructed sources,  we may require 
    new models specifically designed for 3D data cubes, in which the S/N of 
    each channel is typically much lower than for integrated moment maps 
    and valuable information can be extracted from the spectral dimension. 
    The large set of realistic mock data cubes needed for training in this 
    case will be more difficult to produce than training data needed to 
    support the reconstruction of 2D images.
    
	In this paper we introduce a new nonparametric method for reconstructing 
	integral field spectroscopy of lensed galaxies that addresses the above 
	issues. This method is flexible in that it only assumes that the source 
	kinematics are ordered, but does not require that the emitting gas follow 
	any particular spatial or velocity distribution. The approach is a direct 
	extension of the existing 2D nonparametric Bayesian formalism, adapted 
	to fitting all the channels of a data cube simultaneously. In order to 
	appropriately model a source, we introduce a new regularization 
	scheme that is better suited to the form of the data and the properties of 
	the source. 

	This paper is organized as follows. In Section, \ref{sec:bayesian_framework} 
	we present the modified source reconstruction formalism. We describe the 
	mock observations in Section \ref{sec:mock_observations}. In Section 
	\ref{sec:discussion}, we compare reconstructions produced with our new 
	method to the existing 2D nonparametric approach.
	Section \ref{sec:conclusions} provides a summary and discusses future work.

	\section{Bayesian Framework}\label{sec:bayesian_framework}
	The Bayesian framework for reconstructing lensed sources allows priors to be placed on the source parameters using a
	regularization term that disfavors unphysical models that overfit the data, while still maintaining a flexible, 
	non-parametric description of the source. Our new method for reconstructing sources with resolved kinematics 
	modifies the existing Bayesian framework 
	by treating the set of channel images as a single source in the three dimensions of the data cube, rather than 
	modeling the channel images as a set of independent observations. This approach is implemented by reformulating the 
	prior on the source parameters (i.e., the regularization term) to relate the channel images in a physically 
	realistic way. Currently used priors generally restrict the smoothness of the reconstructed source in each channel 
	image independently. The method detailed below instead enforces a restriction on the smoothness of the entire 3D 
	surface brightness distribution in the data cube. We refer to approaches that treat the channel images as 
	independent as 2D regularization methods, and approaches that model all of the channels of the cube together, such 
	as the one introduced here, as 3D regularization methods.
	
	In this section, we summarize the existing 2D regularization approach for single images, its generalization to 
	multiple images, and finally our changes to the formalism to incorporate all of the spectral information available in 
	the data. The general Bayesian formalism is detailed in \citet{Warren2003}, \citet{Suyu2006}, and \citet{Vegetti2009}, 
	while the details of our particular (2D) implementation are discussed in \citet{Tagore2014} and \citet{Tagore2016}.
	
	\subsection{Existing 2D Bayesian Framework}
	First, we briefly review how the Bayesian formalism operates when a single lensed image is reconstructed. Let the 
	data be represented as a 1D vector $\data$ with components $d_k$, where $k = 1, ..., N_d$ and 
	$N_d$ is the number of data points. The data vector can be constructed by concatenating rows of pixels from the 
	image into a 1D vector. The data vector is related to the vector of source parameters 
	$\src$ by
	\begin{align}
	\data = \lens\src + \noise,
	\end{align}
	where $\lens$ is the lensing operator 
	and $\noise$ is the noise in the observed data. The lensing operator maps source parameters to image-plane (data) 
	pixel values and is defined to include the effects of both lensing and blurring due to an observational point-spread 
	function (PSF).
	\footnote{In the case of aperture synthesis data, the synthesized beam serves as the equivalent of a PSF.} The 
	source vector may be represented in multiple ways.  The most straightforward is as a series of 
	surface brightness values on a grid (regular or irregular); however, the surface brightness can also be defined via 
	a set of basis functions \citep[such as shapelets; see][]{Refregier2003,Birrer2015,Tagore2016}, in which case the 
	source parameters become a set of coefficients. We have restricted the scope of this work to grid-based 
	reconstructions, although in principle an approach similar to the one described below can be applied to a 
	shapelet-based source. We note that this formalism can also be applied to visibility-based data, which can be modeled 
	using the same source representations.
	
	\subsubsection{Most Likely Source}
	The most likely source can be obtained by finding the set of source parameters that maximizes the likelihood given by
	\begin{align}\label{eq:likelihood}
	P(\data | \lens, \src) \propto \exp\left(-E_d(\data | \lens, \src)\right),
	\end{align}
	where
	\begin{align}\label{eq:chisq}
	E_d(\data | \lens, \src) = \frac{1}{2} \chi^2(\src) = \frac{1}{2} \left(\lens\src - \data\right)^\top \cov^{-1} \left(\lens\src - \data\right),
	\end{align}
	and $\cov$ is the covariance matrix. \citet{Suyu2006} give the solution for the most likely source vector as
	\begin{align}\label{eq:s_ml}
	\src_{\text{ML}} = \Fmat^{-1} \Dmat,
	\end{align}
	where $\Fmat \equiv \lens^\top \cov^{-1} \lens$ and $\Dmat \equiv \lens^{\top} \cov^{-1} \data$.
	
	\subsubsection{Most Probable Source}
	In the presence of noise, the most likely solution tends to overfit the data, yielding an unphysical source surface 
	brightness distribution. Writing out the full Bayesian posterior probability distribution allows us to place priors 
	on the reconstructed source that effectively smooth the derived source and prevent overfitting. The posterior is given by
	\begin{align} \label{eq:posterior_1}
	P(\src | \data, \lens, \lambda, \reg) \propto \exp\left(-M(\src))\right),
	\end{align}
	where
	\begin{align}\label{eq:M}
	M(\src) \equiv E_d(\src) + \lambda E_s(\src),
	\end{align}
	and
	\begin{align}\label{eq:regularization}
	E_s(\src) = \frac{1}{2}\src^\top \reg \src
	\end{align}
	is a penalty function that enforces the priors that regularize the source parameters, and $\lambda$ is the 
	``regularization strength,'' which controls the relative contributions of the two terms in Equation \ref{eq:M}. 
	When $\lambda$ is large, $E_s(\src)$ prevents the source from overfitting to the data by forcing it to be smooth; 
	when $\lambda$ is small, the source becomes more complex in order to better fit the data. For computational 
	efficiency, the penalty function must be expressible as a quadratic form as in Equation \ref{eq:regularization}, and 
	the regularization matrix $\reg$ must be both symmetric and invertible.
	(This approach to fitting is known generally as Tikhonov regularization \citep{Tikhonov1995} or ridge regression, although
	applications in most other contexts often use $\reg \equiv \Identity$.)
	These conditions restrict our choice of a new regularization matrix in Section 
	\ref{sec:new_framework}.
	
	The matrix $\reg$ is defined according to the specific regularization scheme. Common choices compute the square of 
	the surface brightness gradient or Laplacian summed over all of the pixels, roughly corresponding to the expectation 
	that a realistic surface brightness distribution should be relatively smooth. We emphasize here that the 
	regularization schemes are somewhat arbitrary; most reasonable choices do similarly well at mitigating the 
	overfitting problem but produce differences in the reconstructed sources. For example, zeroth-order 
	($\reg = \Identity$) regularization biases a source toward a ``zero'' (featureless) source; first-order 
	(gradient-based) regularization biases toward a constant-value source; and second-order (Laplacian-based) 
	regularization biases toward a planar source.
	
	The most probable source vector $\src_{\text{MP}}$ maximizes Equation \ref{eq:posterior_1}. For a given value of 
	$\lambda$, the most probable source is given by \cite{Suyu2006} as
	\begin{align}\label{eq:s_mp_2D}
	\src_{\text{MP}} = \Amat^{-1} \Fmat \src_{\text{ML}},
	\end{align}
	where $\Amat = \Fmat + \lambda \reg$. However, to find the best estimate of the source vector, we must 
	find the optimal value of $\lambda$. This step can be done (see \citealt{Suyu2006}) by maximizing the posterior for 
	$\lambda$: 
	\begin{align}\label{eq:lambda_posterior}
	P(\lambda | \data, \lens, \reg) \propto P(\data | \lambda, \lens, \reg) P(\lambda).
	\end{align}
	The prior is typically assumed to be uniform in $\log \lambda$ because the scale of $\lambda$ is not known a 
	priori.
	
	Formally, the most probable source is found by marginalizing over $\lambda$; however, Equation 
	\ref{eq:lambda_posterior} is sharply peaked and can be well approximated by a delta function. As a result, simply 
	using the value of $\lambda$ that maximizes the posterior works well in practice. \cite{Suyu2006} discuss this 
	issue in more detail and give the equation to maximize to find $\lambda$ (see their Equation 19).
	
	\subsubsection{Generalization to Multiple Observations}
	For a set of data vectors $\{\data_i\}$ comprising $N_c$ images, e.g., as obtained in multiple filters or multiple 
	velocity channels, the most general form of Bayes's theorem is
	\begin{align}\label{eq:general_posterior_1}
	P(\{\src_i\} | \{\data_i\}, \lens, \{\lambda_i\}, \reg) \propto & P(\{\data_i\} | \{\src_i\}, \lens) \\
	& \times P\left(\{\src_i\} | \{\lambda_i\}, \reg\right) \nonumber.
	\end{align}
	On the right-hand side, the first term is the likelihood for all of the data, while the second term is a prior over 
	all of the source parameters.
	
	For simplicity, we assume throughout this paper that the data points in each observation correspond to the same sets 
	of pixels (i.e., the pixelization is the same and the same mask is applied to all images) so that each observation 
	uses the same lensing operator $\lens$. It follows straightforwardly from Equation \ref{eq:chisq} that the full 
	likelihood can be factored into a product of likelihoods for individual images or channels:
	\begin{align}\label{eq:likelihood_factored}
	P(\{\data_i\} | \{\src_i\}, \lens) = \prod_{i = 1}^{N_c} P(\data_i | \src_i, \lens).
	\end{align}
	
	The prior can be written in several different ways. The simplest is as a product of factors for the individual 
	images/channels:
	\begin{align}\label{eq:prior_factored_1}
	P\left(\{\src_i\} | \{\lambda_i\}, \reg\right) = \prod_i P\left(\src_i | \lambda_i, \reg\right).
	\end{align}
	This approach is equivalent to treating each of the observations as independent. The most probable set of source 
	channel images $\{\src_i\}$ is then found by maximizing the posterior for each $\lambda_i$ independently. For a set 
	of observations in several photometric filters, this approach is likely the best one in the absence of a model 
	relating the emission across the filters. If the data are closely spaced channels in a data cube from an integral 
	field spectrograph or a radio telescope, however, the emission in consecutive channels is related by the source 
	kinematics, making it non-optimal to treat the channels independently. The next section lays out a more informed way 
	of writing the prior for sources with resolved kinematics.

	\subsection{Updated 3D Bayesian Framework}\label{sec:new_framework}	
	We assume here that (i) our data are a set of channel images from a data cube with uniform angular resolution, and 
	(ii) the channels have the same masks applied, such that each pixel that exists in any channel image exists in every channel 
	image. The latter assumption ensures that we can reconstruct using source plane grids that are identical for all 
	channels. We can then define the vector $\data$ for all of the data points $\{\data_i\}$ as the concatenation of 
	all of the data vectors for each of the channel images. We choose to continue to describe the source as a set of 
	channel images, so that the source vector $\src$ can be defined similarly. (In principle, the source vector can be 
	defined in an alternative way, e.g., using a set of 3D basis functions, but such a scenario is beyond the scope of the 
	present paper.)
	
	Throughout this section, we represent matrices that act on the vector of combined data cube images with a hat, e.g., 
	$\Lhat$. Unless otherwise defined, these are block-diagonal matrices formed by $N_c$ blocks of the original matrix. 
	For example,  $\Lhat \equiv \Identity_{N_c} \otimes \lens$, where $\Identity_{N_c}$ is the $N_c \times N_c$ 
	identity matrix and $\otimes$ denotes the Kronecker product. Following this notation, the likelihood, which can be 
	factored as in Equation \ref{eq:likelihood_factored}, can instead be written as in Equation \ref{eq:likelihood} 
	after replacing un-hatted matrices with their hatted counterparts:
	\begin{align}
	E_d(\data | \src, \Lhat) = \frac{1}{2} \left(\Lhat\src - \data\right)^\top \Chat^{-1} \left(\Lhat\src - \data\right).
	\end{align}
	The block-diagonal form of $\Lhat$ persists as long as adjacent channels are 
	independent, since $\lens$ only includes blurring by the 2D PSF. This 
	assumption holds for radio data, which we consider in this work, but not 
	for some shorter-wavelength data where a line-spread function (LSF) means adjacent channels are not independent. 
	Appendix \ref{app:lsf} discusses how the $\Lhat$ matrix can be changed to 
	account for an LSF in such cases.

	\subsubsection{3D Regularization}\label{sec:3D_regularization}
	As discussed above, the 2D regularization schemes allow us to reconstruct a 
	relatively smooth source in 2D without prescribing a particular 
	emission profile. Nevertheless, the specific choice of regularization matrix 
	influences the properties of the source. A rigorous Bayesian analysis 
	incorporates variation of priors in order to understand the effect of prior 
	choice on the results. This fact motivates us to reformulate our prior 
	(the regularization matrix) in order to better reflect the expected 
	properties of the source in 3D. This matrix should impose a similar spatial 
	smoothness constraint on the reconstruction while also requiring that 
	emission change smoothly from one channel to the next. We do this by 
	defining a new regularization matrix that penalizes the squared 3D gradient 
	(or Laplacian) of the source summed over all pixels in the source data cube.
	
	In contrast to Equation \ref{eq:prior_factored_1}, we write the prior for the 3D regularization in the form
	\begin{align}
	P(\{\src_i\} | \{\lambda_i\}, \reg) & = P(\src| \RthreeDhat(\lambda, \eta)) \nonumber \\
	& \propto \exp\left(-\frac{1}{2} \src^\top \RthreeDhat \src\right),
	\end{align}
	where $\RthreeDhat(\lambda,\eta)$ is the new 3D regularization matrix and we use only two regularization strengths, 
	$\lambda$ and $\eta$. Note that we have absorbed $\lambda$ and $\eta$ into the regularization matrix, which we 
	define below.
	
	The 2D regularization matrix for a single channel image is given by $\RtwoD = \HtwoD^\top \HtwoD$, where $\HtwoD$ is 
	the matrix that computes the components of the gradient (or Laplacian) for all of the pixels in the source. 
	Similarly, the 2D regularization matrix for a set of channel images is given by 
	$\RtwoDhat \equiv \HtwoDhat^\top \HtwoDhat$, where $\HtwoDhat = \Identity_{N_c} \otimes \HtwoD$. We can similarly 
	define the matrix $\HthreeDhat$ that computes the components of the gradient (or Laplacian) in 3D as
	\begin{align}
	\HthreeDhat \equiv \lambda^{1/2} \HtwoDhat + \eta^{1/2} \Qhat,
	\end{align}
	where $\Qhat$ contains the extra terms that specify the components of the gradient or Laplacian in the third 
	dimension. Appendix \ref{app:Qhat} provides a derivation of $\Qhat$ for both of these cases. The 3D regularization 
	matrix is then
	\begin{align}
	\RthreeDhat & \equiv \HthreeDhat^\top \HthreeDhat \nonumber \\
	& = \lambda \RtwoDhat + \eta \Qhat^\top \Qhat \nonumber \\
	& \quad + \lambda^{1/2} \eta^{1/2} \left(\HtwoDhat^\top\Qhat + \Qhat^\top\HtwoDhat\right).
	\end{align} 
	When $\eta = 0$, this regularization scheme reduces to the 2D case with $\lambda$ identical for all channels.
	
	We have introduced a second regularization strength, $\eta$, to control the smoothness of the reconstruction in the 
	spectral direction, analogous to how $\lambda$ controls its spatial smoothness. We refer to $\lambda$ and $\eta$ as 
	the spatial and spectral regularization strengths, respectively. It is necessary to have {\it two} such strengths 
	because the spatial and spectral dimensions of the cube have different units and independent smoothness requirements.
		
	In the 2D context, it is well understood that regularization affects the properties of the reconstructed source. The 
	addition of the regularization term results in a biased estimator of the real source, in the sense that the 
	reconstruction mitigates the effect of the noise in the data by introducing correlations between neighboring 
	source-plane pixels. The scale over which this spatial correlation is present is determined by the regularization 
	strength, such that larger values of $\lambda$ cause correlations over larger scales. Likewise, the 3D regularized 
	source is also a biased estimator, but with correlations now present in both the spatial and spectral directions, on 
	scales determined by the two regularization strengths. We will see in Section \ref{sec:performance} how this bias 
	affects the reconstructed sources.

	\subsubsection{Most Probable Source}
	For given values of $\lambda$ and $\eta$, the most likely source, $\src_{\text{ML}}$, is given by Equation 
	\ref{eq:s_ml} with the matrices replaced by their hatted versions. Likewise, the most probable source is given, 
	analogous to Equation \ref{eq:s_mp_2D}, by
	\begin{align}\label{eq:s_mp_3D}
	\src_{\text{MP}} = \Ahat^{-1} \Fhat \src_{\text{ML}},
	\end{align}
	where we now have $\Ahat = \Fhat + \RthreeDhat(\lambda,\eta)$
	
	\subsubsection{Finding $\lambda$ and $\eta$}
	As in the 2D regularization case, the joint posterior for $\lambda$ and $\eta$ is given by
	\begin{align}\label{eq:lambda_eta_posterior}
	P(\lambda,\eta | \data, \Lhat, \RthreeDhat) \propto P(\data | \lambda, \eta, \Lhat, \RthreeDhat) P(\lambda, \eta).
	\end{align}
	In order to compute the posterior, we must place a prior on the regularization strengths, 
	$P(\lambda, \eta)$. As in the 2D case, we do not know the scale for either regularization strength a 
	priori, and it is natural to assume that both $\lambda$ and $\eta$ have log-uniform priors. Consequently, maximizing 
	Equation \ref{eq:lambda_eta_posterior} is equivalent to maximizing 
	$\ln P(\data | \lambda, \eta, \Lhat, \RthreeDhat)$.
	
	Following the same steps as \citet{Suyu2006}, we can derive the joint posterior for the regularization strengths as
	\begin{align}\label{eq:ln_P_lambda_eta}
	\ln P(\lambda,\eta | \data) & = \frac{1}{2} \ln\abs{\RthreeDhat} - \frac{1}{2} \ln\abs{\Ahat} \nonumber \\
	& \quad + \frac{1}{2} \src_{\text{MP}} \Ahat \src_{\text{MP}} - \frac{1}{2} \ln\abs{\Chat} \nonumber \\
	& \quad - \frac{N_d N_c}{2} \ln 2\pi.
	\end{align}
	This equation is analogous to Equation 19 in \citet{Suyu2006}, but with the vectors and matrices replaced with their 
	larger, hatted versions.
	
	As with the 2D approach, we can maximize Equation \ref{eq:ln_P_lambda_eta} to find the optimal pair of 
	regularization strengths. Again, we want to marginalize over the regularization strengths; however, we still expect 
	the posteriors to be sharply peaked enough that we can use the modes in place of the marginalized distributions. We 
	explore this assumption in Section \ref{sec:reg_strength_optimization}.

	\section{Mock Observations}\label{sec:mock_observations}
	To compare the performance of our new approach to that of existing methods, 
	we create a series of mock observations of carbon monoxide (CO) line 
	emission from galaxy disks at redshift $z \sim 2$--$3$, as would be observed by 
	the Atacama Large Millimeter/submillimeter Array (ALMA). Since our source 
	representation is flexible, we expect that the results of the tests 
	discussed below are not strongly dependent on the chosen parameters of the 
	disk models used to generate the mock observations.
	
	Briefly, the mock data cubes are created as follows. The unlensed source cubes are produced using the Python package 
	GalPaK$^{\text{3D}}$ \citep{Bouche2015}, which generates data cubes from parameterized models of rotating disks. We 
	then use the lens modeling software \code{lensmodel} \citep{Keeton2011} to lens the source channel images. Finally, 
	we convolve the channel images with a synthesized beam and add noise to simulate mock observations. The following 
	subsections describe these steps in more detail; as discussed in Section \ref{sec:discussion}, we have especially 
	focused on exploring the effects of the spectral and spatial resolutions and the S/N ratio of the observations.

	\subsection{Input Source Models}\label{sec:input_models}
	GalPaK$^{\text{3D}}$ allows the functional forms of several components of a disk model to be changed: the galaxy 
	flux profile, disk thickness profile, rotation curve shape, and dispersion profile. 
	Because our lens modeling approach is non-parametric, we expect choices among these options to have little effect 
	on the quality of our reconstructions. Therefore, we leave them fixed at their defaults: an exponential flux 
	profile, Gaussian thickness profile, $\arctan$ rotation curve, and ``thick" dispersion profile. 
	
	Each source cube used as input to \code{lensmodel} is defined by its dimensions, its spatial 
	and spectral resolutions, and the 10 parameters of a disk model (positional center, $x_0$, $y_0$; reference 
	velocity, $v_{z_0}$; half-light radius, $R_{1/2}$; inclination angle, $\theta_{\text{inc}}$; position angle, 
	$\text{PA}$; turnover radius, $r_t$; maximum velocity, $V_{\text{max}}$; velocity dispersion, $\sigma_0$; and 
	overall normalization). The pixel scale of the input source is chosen to be small enough that it does not affect the 
	pixel values in the final lensed image. The final spectral resolution is set when the mock cube is created, while 
	the final spatial resolution is set after the mock images are lensed.

	\subsection{Simulated Observations}\label{sec:simulated_observations}
	\begin{figure*}
		\centering
		\includegraphics[width=5in]{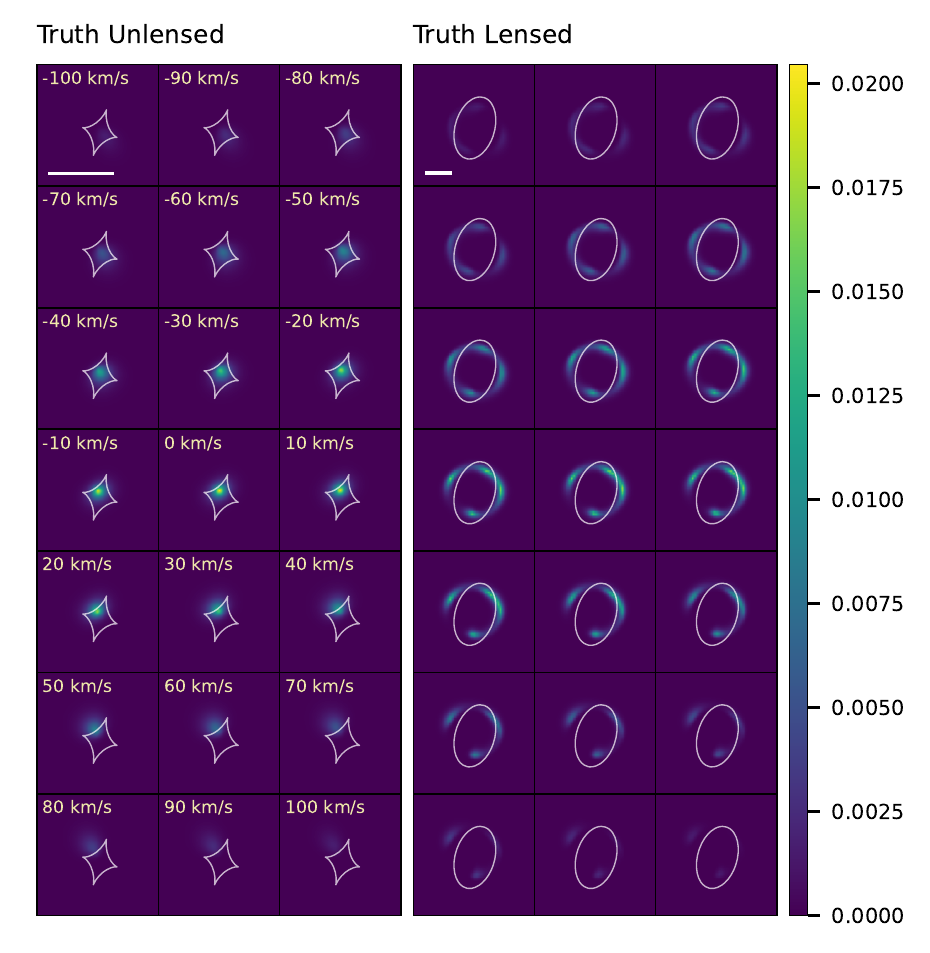}
		\caption{Example of the channel images for the ``true'' unlensed (left) and lensed (right) source used to 
		make the mock observations. The channel width for this example is $10 \,\text{km}\,\text{s}^{-1}$. The lensed 
		images are shown with no noise and no convolution with a synthesized beam. The scale bars show $1''$. The critical curve and caustic are shown in white. The velocity of each channel is shown in the upper left. The color scale shows arbitrary surface brightness units.}\label{fig:truth_images}
	\end{figure*}
	
	After creating a data cube of the unlensed source, we use \code{lensmodel} to lens each of the channel images, choosing 
	lens model parameters that are typical of observed lensed systems. For most examples in this paper, we use as a 
	fiducial lens model a singular isothermal ellipsoid with an Einstein radius $b = 1''$, ellipticity $e = 0.38$, 
	position angle $-16^\circ$, and no external shear. The image plane channel maps (before convolution) are created by 
	interpolating surface brightness values between source-plane pixels. The pixel scale of the lensed channel images is 
	chosen to achieve the desired resolution when convolved with a circular Gaussian beam with a FWHM of 5 pixels. After 
	convolving with the beam, we add uniform Gaussian noise across all channels to achieve a desired S/N for the mock 
	observations.
	
	It should be noted that the position of the source and the orientation of its velocity gradient with respect to the 
	caustic can be expected to have some effect on the comparison between regularization approaches. For a source--lens 
	configuration in which the velocity gradient crosses or closely approaches the caustic, the source channel images 
	will effectively cover a larger number of very different source--lens configurations, providing many constraints on 
	the lens potential. This effect is illustrated in Figure \ref{fig:truth_images}, which shows how the unlensed 
	channel images are lensed into a series of different configurations because of the large source velocity gradient.

	The fiducial data set used throughout this paper is created using the lens 
	model given above and a source model with parameters chosen to be typical 
	for high-redshift galaxies: $R_{1/2} = 1.6 \,\text{kpc} \,(0.2'')$, 
	$\theta_{\text{inc}} = 45^\circ$, $r_t = 1.6 \,\text{kpc} \,(0.2'')$, $V_{\text{max}} = 150 \,\kms$, 
	and $\sigma_0 = 30 \,\kms$. This model corresponds to a dynamical mass of 
	$M_{\text{dyn}} = V_{\text{max}}^2 R_{1/2} / G = 8.4 \times 10^9 \,M_{\odot}$. 
	The final data cube has $10 \,\kms$ channels, a $0.5''$ FWHM beam, and a 
	peak S/N of 30. These channel images are shown in Figure 
	\ref{fig:truth_images} before convolution with the beam and inclusion of 
	noise.

	\section{Comparison between Approaches}\label{sec:discussion}
	\subsection{Implementation}\label{sec:implementation}
	We modify the source reconstruction package \code{pixsrc} \citep{Tagore2014, Tagore2016}, which is an extension to 
	the \code{lensmodel} code \citep{Keeton2011}, to perform the lens modeling and source reconstruction described in 
	Section \ref{sec:bayesian_framework} above. We compare the performance of the 2D and 3D regularization approaches 
	below but first discuss some aspects of the computations here.

	In the tests throughout this paper we use a grid that is constructed by 
	ray-tracing every third data pixel back to the source plane. For our 
	$51 \times 51$ pixel data images, these settings with the fiducial lens 
	model result in a total of $873$ pixels to describe one image of the source. 
	Our fiducial $21$ channel dataset thus requires a total of $18,333$ free 
	parameters to describe the entire source-plane cube. Source-plane images 
	throughout this paper are produced by interpolating the irregular source-plane 
	pixel values onto a grid of smaller square ``visualization pixels.'' 
	The size of the visualization pixels is the \code{pixsrc} default of $1/4$ 
	the data pixel size.
	
	For ease of comparison with the bulk of the lensing literature, we use Laplacian-based regularization.
	The $\RthreeDhat$ matrix described above is very sparse for data with even a few channels. The $\HtwoDhat$ matrix is 
	block-diagonal and built up from copies of $\HtwoD$, which itself is sparse, and $\Qhat$ has only three non-zero 
	diagonals (see Appendix \ref{app:Qhat}). As a result, we can efficiently solve matrix equations and compute 
	determinants using existing sparse matrix libraries that can take advantage of multi-threaded processors. We use \code{scikit-sparse}\footnote{https://github.com/scikit-sparse/scikit-sparse}, which provides a python interface to the CHOLMOD library \citep{CHOLMOD}, for sparse matrix calculations.
	Nevertheless, the matrices are still quite large ($\Ahat$ is $N_s N_c \times N_s N_c$), and computation times may be a 
	potential concern for datasets with a very large number of channels. 
	
	Finding the optimal regularization strength consumes a significant fraction of the computation time even when using 
	2D regularization. Because of the large matrices and the fact that we are now searching for \textit{two} optimal 
	regularization strengths, identifying a local minimum becomes more of a problem for 3D regularization. 
	We have not yet explored methods for speeding up this parameter search.
	
	In the tests below, we fix the lens model to the true one used in generating the mock observations. The likely 
	approach for modeling real data would be to derive an initial lens model from the zeroth-moment map and use this as 
	a starting point for further modeling of the full data cube. This two-step approach should cut down significantly on 
	computation time compared to a full lens model optimization with 3D regularization. We defer to a future paper an 
	investigation of the effects of our approach on the final derived lens model parameters and corresponding source 
	intensity distribution.

	\subsection{Behavior and Performance}\label{sec:performance}
	We perform most of our comparisons using the fiducial source and lens models described in Section 
	\ref{sec:simulated_observations} for a mock dataset with $0.5''$ spatial resolution, 
	$10 \,\textrm{km}\,\textrm{s}^{-1}$ spectral resolution, and a peak S/N radio of 30. We vary these observational 
	parameters in Sections \ref{sec:spectral_resolution} and \ref{sec:spatial_resolution} to assess their effects on the 
	reconstruction properties.
	
	\subsubsection{Optimization of regularization strengths}\label{sec:reg_strength_optimization}
	\begin{figure}
		\centering
		\includegraphics[width=3.35in]{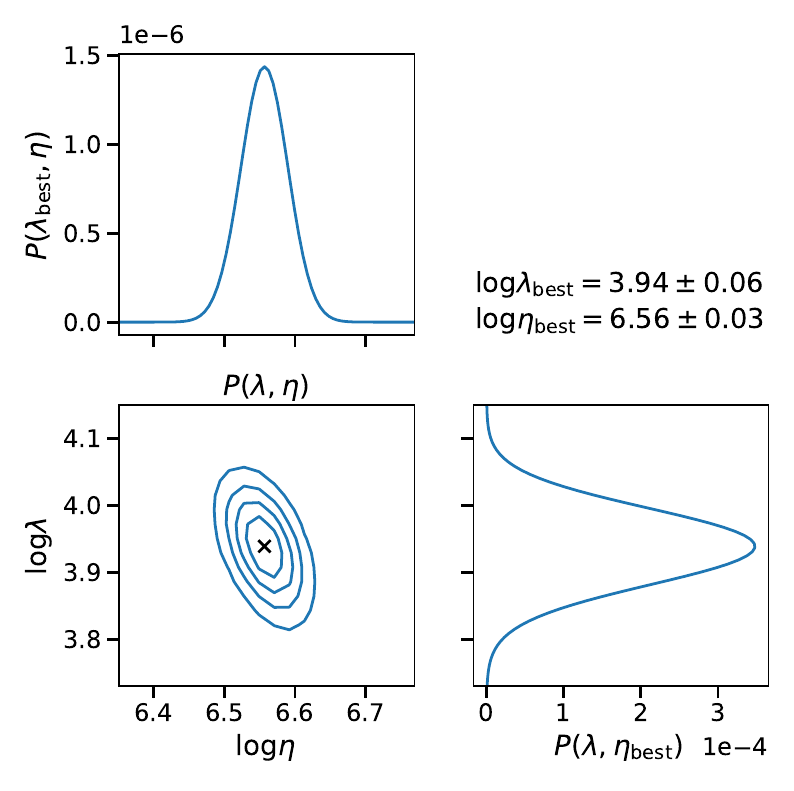}
		\caption{
			Views of the regularization strength posterior probability distribution. Lower left: full posterior 
			distribution $P(\lambda,\eta)$ with the peak from the numerical optimization, 
			$(\lambda_{\rm best},\eta_{\rm best})$, marked with a black X. The contour levels are 20\%, 40\%, 60\%, and 80\% 
			of the peak. The two conditional posteriors are shown, (above) $P(\lambda,\eta_{\rm best})$ and (right) 
			$P(\lambda_{\rm best},\eta)$. Note the small ranges of regularization strengths.}\label{fig:f_lambda_eta}
	\end{figure}
	
	The optimal regularization strengths can be found by optimizing the posterior given in Equation 
	\ref{eq:ln_P_lambda_eta}. In order to account for correlations between pixels in the images, we use the noise 
	scaling procedure described in \citet{Riechers2008}. Without noise scaling, the reconstructed sources produce 
	residuals with an RMS significantly lower than the data noise level. To produce an appropriately smooth source, we 
	scale the input noise level by a factor that results in residuals with the expected RMS value. Because this noise 
	scale factor should be dependent on properties of the data, we determine its value using 3D regularization and fix 
	it for all other regularization schemes with the same data.
	
	As discussed above, we expect that the 3D regularization strength posterior should be narrow as in the 2D case 
	\citep{Suyu2006}, allowing a simple optimization of the regularization strengths to 
	determine the best source. We test this expectation by plotting the posterior for the two 3D regularization 
	strengths in Figure \ref{fig:f_lambda_eta}. The lower left panel shows the joint posterior $P(\lambda,\eta)$ for the 
	fiducial dataset, while the upper left and lower right panels show the conditional posteriors 
	$P(\lambda,\eta_{\rm best})$ and $P(\lambda_{\rm best},\eta)$, respectively. The optimal values of the 
	regularization strengths were found to be $(\log\lambda_{\rm best},\log\eta_{\rm best}) = (3.94,6.56)$, and the 
	standard deviations of the conditional posteriors were $(\sigma_{\log\lambda},\sigma_{\log\eta}) = (0.06,0.03)$.
	We then reconstruct sources two standard 
	deviations (in log-space) from the peak on either side of each conditional 
	posterior distribution with the other 
	regularization strength fixed to the best value. We find that these sources 
	deviate by no more than $\sim 3\%$ of the 
	peak of the best source. This indicates that the posterior is narrow 
	compared with the scale over which the reconstructed source changes
	appreciably, such that we can approximate sources near the peak by the best 
	source.

	\subsubsection{General Reconstruction Properties}\label{sec:reconstructed_source_properties}
	\begin{figure*}
		\centering
		\includegraphics[width=\textwidth]{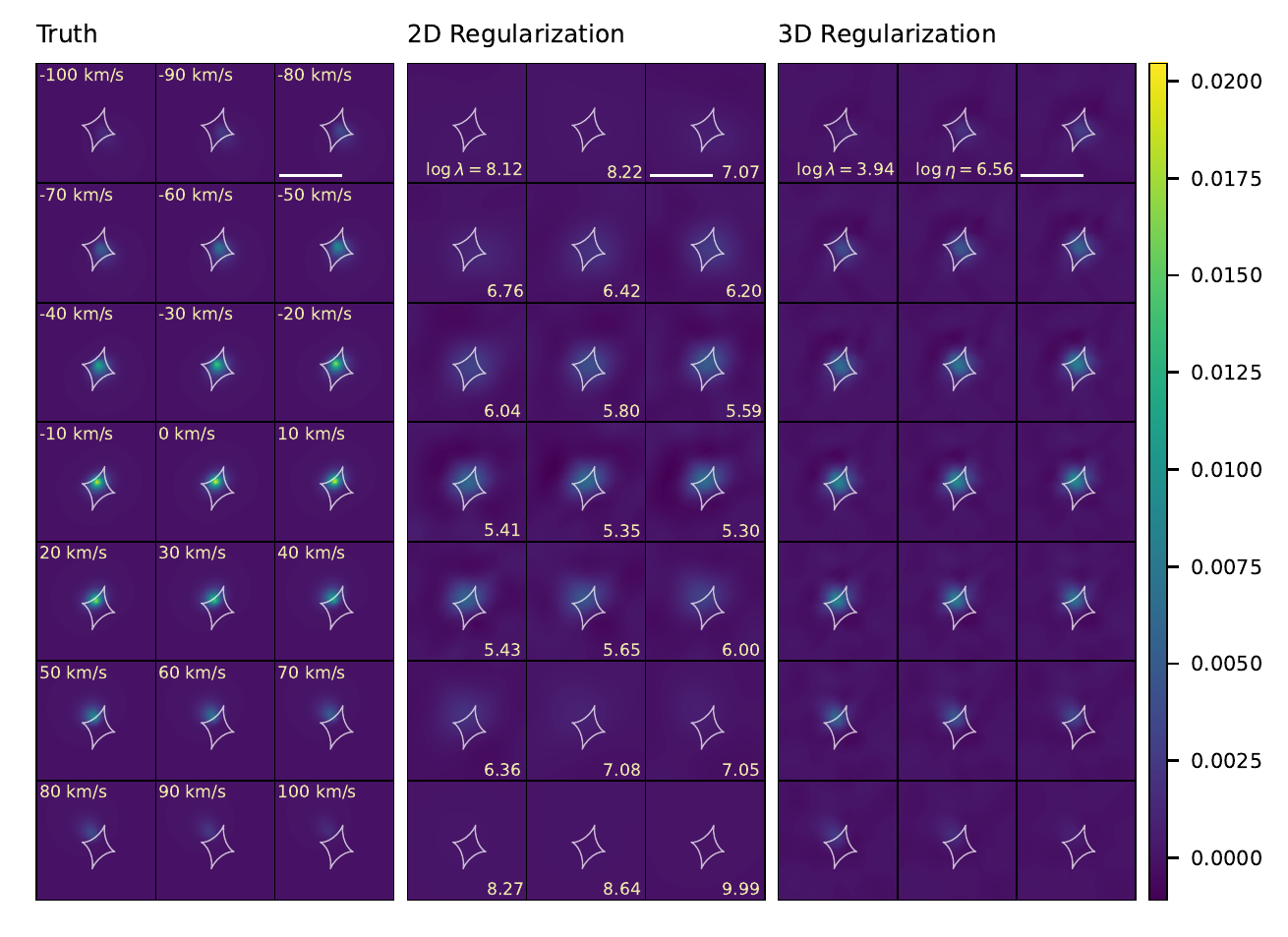}
		\caption{Comparison of the different source reconstruction techniques for the fiducial mock data cube with 
		channel width $10 \,\text{km}\,\text{s}^{-1}$. From left to right: the ``true" (input) source, the 
		reconstruction using 2D regularization, and the reconstruction using 3D regularization. The scale bars show 
		$1''$. The caustic is shown in white. The velocity of each channel is shown at the upper left. The 2D 
		regularization strength is shown in the lower right of each relevant panel, and the 3D regularization 
		strengths are shown in the first two panels for the 3D regularized source. The 3D reconstructed sources are 
		significantly more compact than their 2D counterparts and are detected better in the outer channels where the 2D 
		reconstructed sources are not.}\label{fig:source_truth_comparison}
	\end{figure*}
	
	\pagebreak
	\begin{figure*}
        \rotatebox{90}{
            \begin{minipage}[c][\textwidth][c]{\textheight}
        		\centering
        		\includegraphics[width=\textwidth]{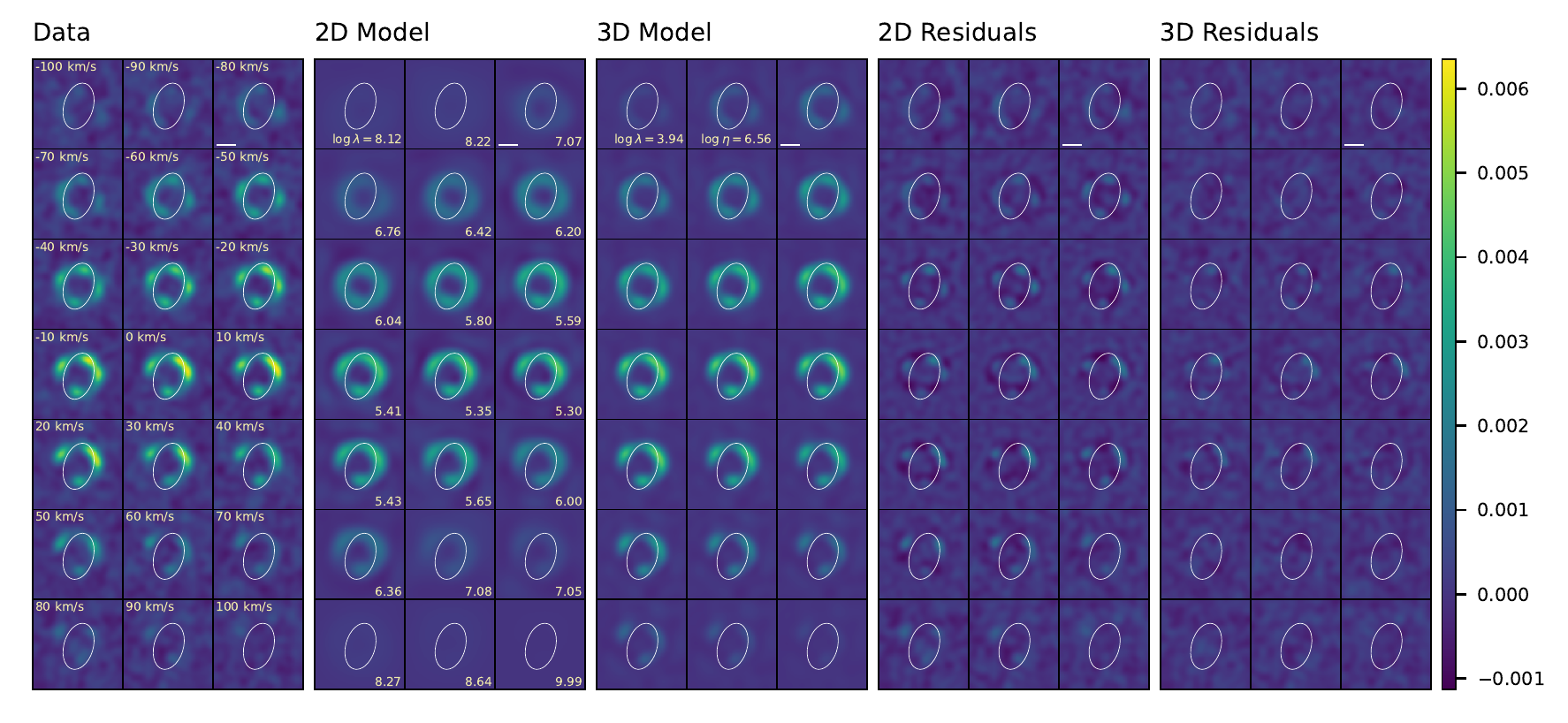}
        		\caption{Comparison of the source models in the image plane. From left to right: the simulated data, 
				the lensed 2D regularization model, the lensed 3D regularization model, the 2D regularization residuals, 
				and the 3D regularization residuals. The residuals show that the 3D regularization gives a better fit to the observed data with residuals much closer to the noise level. The critical curve is shown in white; regularization strengths are indicated as in Figure \ref{fig:source_truth_comparison}.}\label{fig:image_plane_comparison}
    		\end{minipage}
		}
	\end{figure*}

	\begin{figure*}
		\centering
		\includegraphics[width=6in]{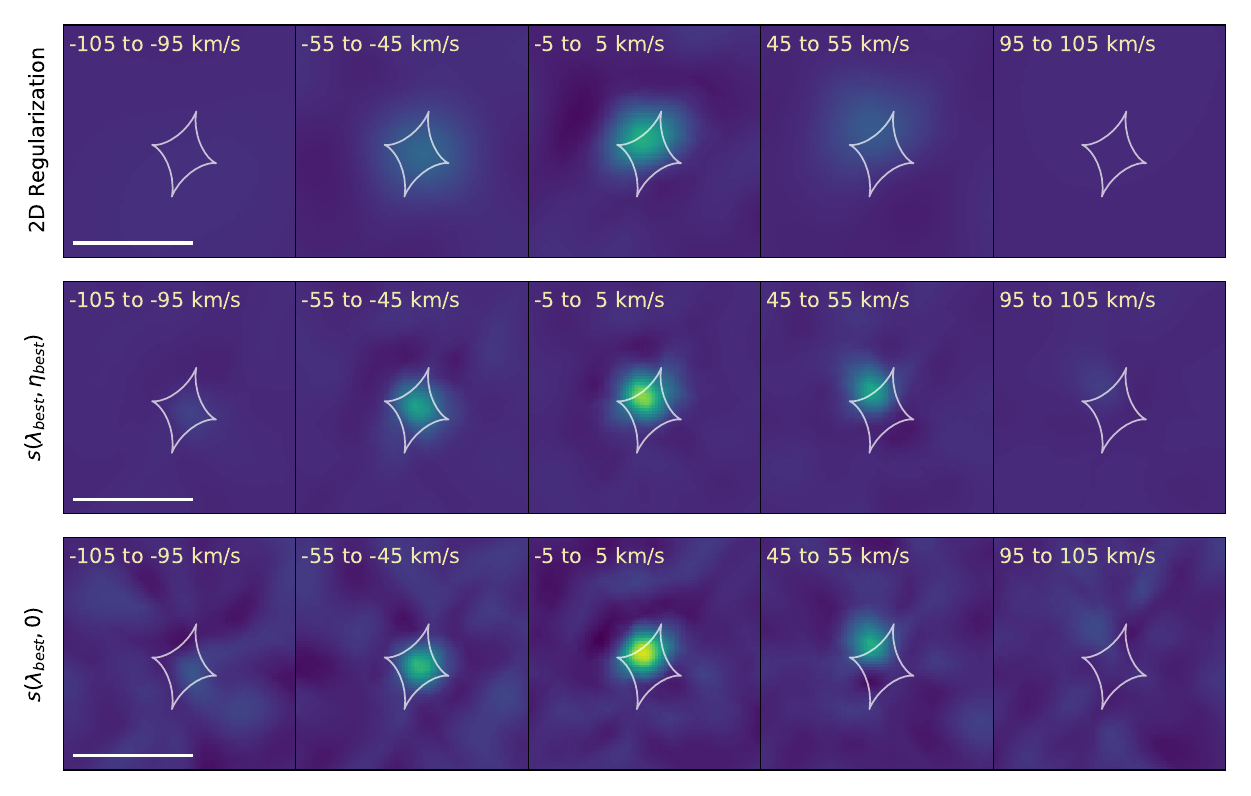}
		\caption{Comparison between sources reconstructed with different regularization schemes. Rows (top to bottom): 
		the 2D-regularized reconstruction with regularization strengths given in Figure \ref{fig:source_truth_comparison}, the 3D-regularized reconstruction with optimal regularization strengths 
		$(\lambda_{\text{best}},\eta_{\text{best}}) = (10^{3.94}, 10^{6.56})$, and the 3D-regularized reconstruction with regularization 
		strengths $(\lambda_{\text{best}},0)$. Each row shows five channel images selected from the reconstructed cube. 
		The higher noise in the bottom row compared to the middle row shows how the spectral portion of the 3D 
		regularization (i.e., the terms in $\Qhat$) suppresses noise in the spectral 
		direction.}\label{fig:compare_eta_effect_subset}
	\end{figure*}
	
	\begin{figure*}
		\centering
		\includegraphics[width=6in]{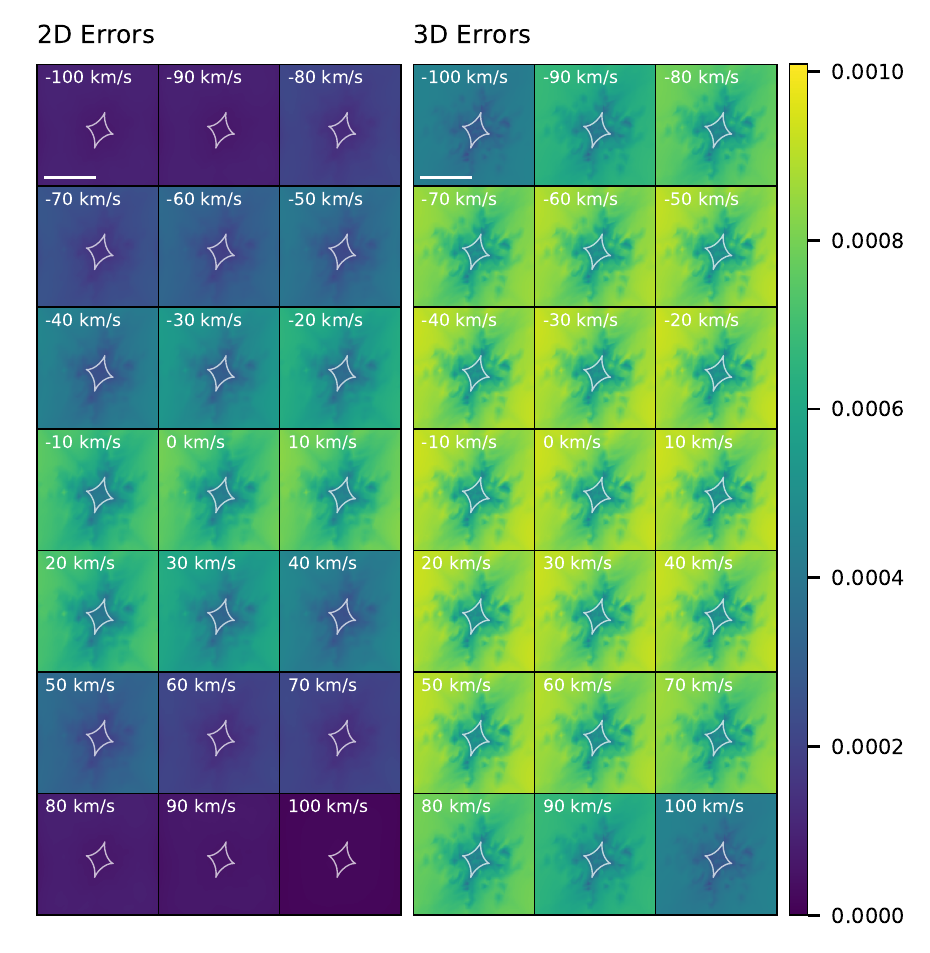}
		\caption{Source plane errors for 2D and 3D regularization. For 2D regularization, the magnitude and spatial 
		structure of the errors depend on the value of the spatial regularization strength, $\lambda$, for each channel. 
		For the 3D regularization, the magnitude and spatial structure of the errors is more uniform across channels 
		(except for edge channel effects) due to the structure of the 3D regularization matrix.}\label{fig:source_errors}
	\end{figure*}
	
	\begin{figure*}
		\centering
		\includegraphics[width=6in]{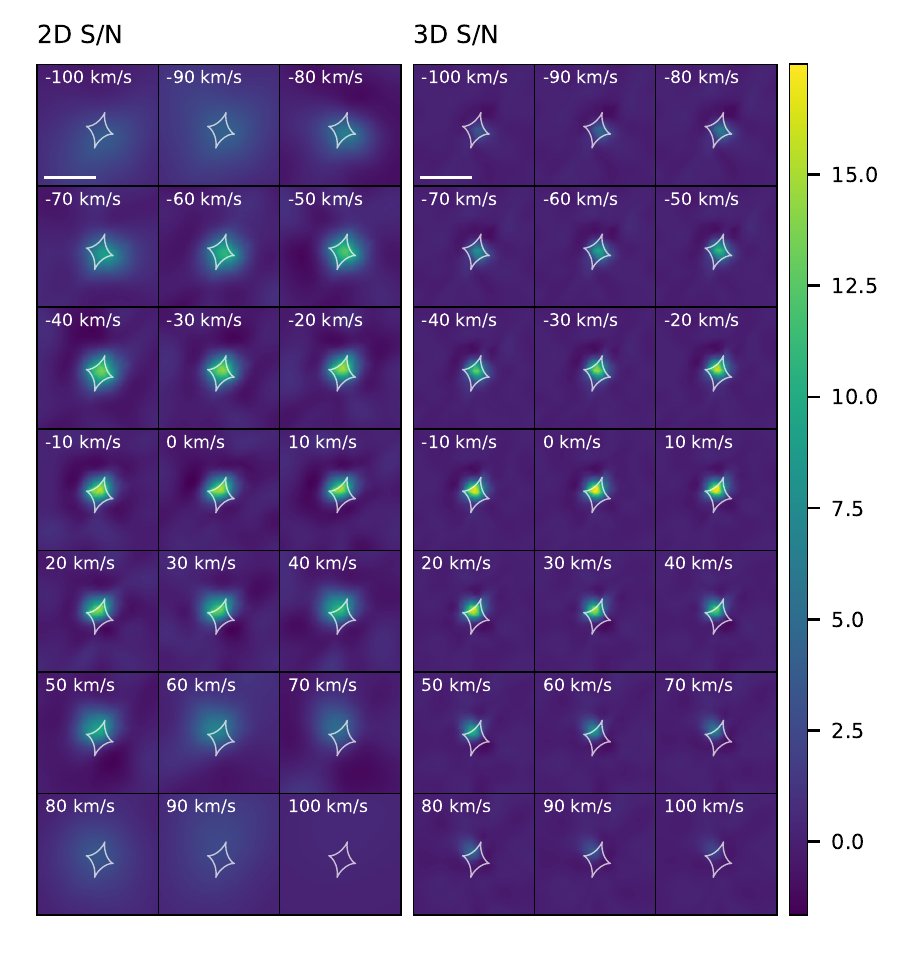}
		\caption{Maps of the source plane S/N ratio for 2D and 3D regularization. Despite the generally higher noise 
		level of the 3D regularization shown in Figure \ref{fig:source_errors}, the peak S/N ratio is higher because of 
		the higher effective resolution.}\label{fig:source_s2n}
	\end{figure*}

	Reconstructions of the fiducial dataset using both regularization schemes are compared to the true input source in 
	Figure \ref{fig:source_truth_comparison}. Figure \ref{fig:image_plane_comparison} shows the mock data that are used 
	for the reconstructions as well as the lensed and PSF blurred model images and their respective residuals. We find 
	that the 3D regularization consistently produces source-plane channel maps that are an overall better match to the 
	data and the true source. The source reconstructed with the new method is more compact than its 
	2D counterpart, with a higher peak surface 
	brightness and an overall profile that more closely resembles that of the truth. Importantly, the 3D reconstruction 
	also better reproduces the emission in the low-S/N outer channels where the 2D reconstruction fails. 
	The RMS of the difference between the truth and the reconstructions is 50\% higher for the 2D 
	regularization, at 0.00031 vs. 0.00020 for the 3D regularization in arbitrary surface brightness units. (These numbers are also consistent with the error images shown in Figure \ref{fig:source_errors} below.) 

	The more compact 3D reconstruction is the result of a smaller effective source plane beam (mostly the consequence 
	of a smaller spatial regularization strength $\lambda$), although the effect of the new regularization scheme on 
	the resolution and noise properties of the reconstructions is more complex. Both schemes smooth over noise and 
	``ringing'' in the maximum likelihood reconstructions by creating correlations, at the cost of 
	a lower effective resolution. For the 3D regularization, the smoothing in the spectral direction comes from the 
	$\eta$-terms in $\RthreeDhat$. This effect can be seen in Figure \ref{fig:compare_eta_effect_subset}, which shows 
	selected channels of the best 2D reconstruction, the best 3D reconstruction with 
	$(\lambda,\eta)=(\lambda_{\text{best}},\eta_{\text{best}})$, and the 3D reconstruction with 
	$(\lambda,\eta)=(\lambda_{\text{best}},0)$. The last of these shows how significantly the noise is smoothed across 
	channel images.

	Another way to view the better performance of the 3D regularization is to 
	note the number of free parameters in each of the regularization schemes. 
	The 2D regularization has $N_c = 21$ regularization strengths, while 
	the 3D regularization has only two. Despite this, the 3D regularization 
	achieves a better fit because the regularization matrix is better suited to 
	the data.
	
	Because of the structure of the regularization matrices, 2D 
	regularization produces correlations in only the spatial directions, while 
	3D regularization also does so in the spectral direction. 
	As a result of this behavior, we must now discuss the resolution in terms 
	of a 3D resolution element. For both schemes, the resolution 
	element associated with each pixel changes as a function of position due to 
	the variation of the magnification factor across the source plane.
	
 	We do not provide a full quantification of the source-plane 
	resolutions in this work, but we discuss some of the issues with the resolution analysis and provide a visualization 
	of the differences between the source-plane resolutions for both reconstruction methods in Appendix 
	\ref{app:resolution}. In short, the 2D resolution element is effectively the width of a single channel in the 
	spectral direction, but has varying effective beam 
	sizes depending on the S/N of the channel data. The most compact of these beams (near systemic velocity, where S/N 
	is highest) have sizes similar to those delivered by the 3D regularization but show more prominent spatial ringing 
	features, while the outer channels have beams so 
	large that the emission cannot be located reliably. In contrast, the 3D regularization beam is more	compact 
	spatially and identical across all channels, with reduced spatial ringing. The 3D regularization beam also spreads emission spectrally across a $\sim \pm 20 \,\text{km}\,\text{s}^{-1}$ range. This effect occurs because of the structure 
	of $\RthreeDhat$, which correlates emission between neighboring channels via the $\eta$-terms but enforces the same 
	spatial component of the beam by only using one spatial regularization strength $\lambda$.

	We can also look at the source plane noise properties of the reconstructed 
	cubes, which can be obtained from the diagonal of the source covariance 
	matrix $\bm{C}_s = \Ahat^{-1}$ \citep{Suyu2006}. Figure 
	\ref{fig:source_errors} shows the channel-by-channel 
	noise maps across the source plane for both reconstruction methods. Similar to the beam properties, the 3D 
	regularization yields more uniform noise properties, with the overall amplitude of the noise maps remaining constant 
	across nearly all channels (except for the first and last few channels, which have lower overall normalizations due 
	to edge effects). The 2D regularization noise level varies dramatically depending on each channel's 
	$\lambda_i$, with the very low outer-channel noise amplitudes a direct result of the 
	coarse spatial resolution needed to reconstruct those channels. The overall noise level appears higher for the 
	3D relative to the 2D regularization, although most of this increase occurs away from the caustic. In the central 
	high-S/N channels, the 3D regularization shows only a modest increase in noise relative to the 2D maps. In the outer 
	channels, the increased resolution delivered by the 3D regularization is the main reason for the large difference 
	between the two methods.

	We note as a caveat that Figure \ref{fig:source_errors} does not display the covariances between pixels. These 
	covariances may be important for understanding how the noise is correlated between nearby channels in the 
	reconstructed 3D cube. While this issue may require further exploration, the comparison between $0^{\textrm{th}}$ 
	moment maps in Section \ref{sec:moment_maps} below suggests that these correlations are not significant.

	Because the 3D regularization achieves a smaller effective beam size, we expect that the amplitudes of both the 
	emission and noise will be increased. To visualize this effect, we plot in Figure \ref{fig:source_s2n} the S/N 
	of the reconstructions for each channel. The peak S/N of the cubes is $15.5$ and $17.5$ for the 2D and 
	3D reconstructions, respectively.

	\subsubsection{Source Moment Maps}\label{sec:moment_maps}
	\begin{figure*}
		\centering
        \includegraphics[width=7.0in]{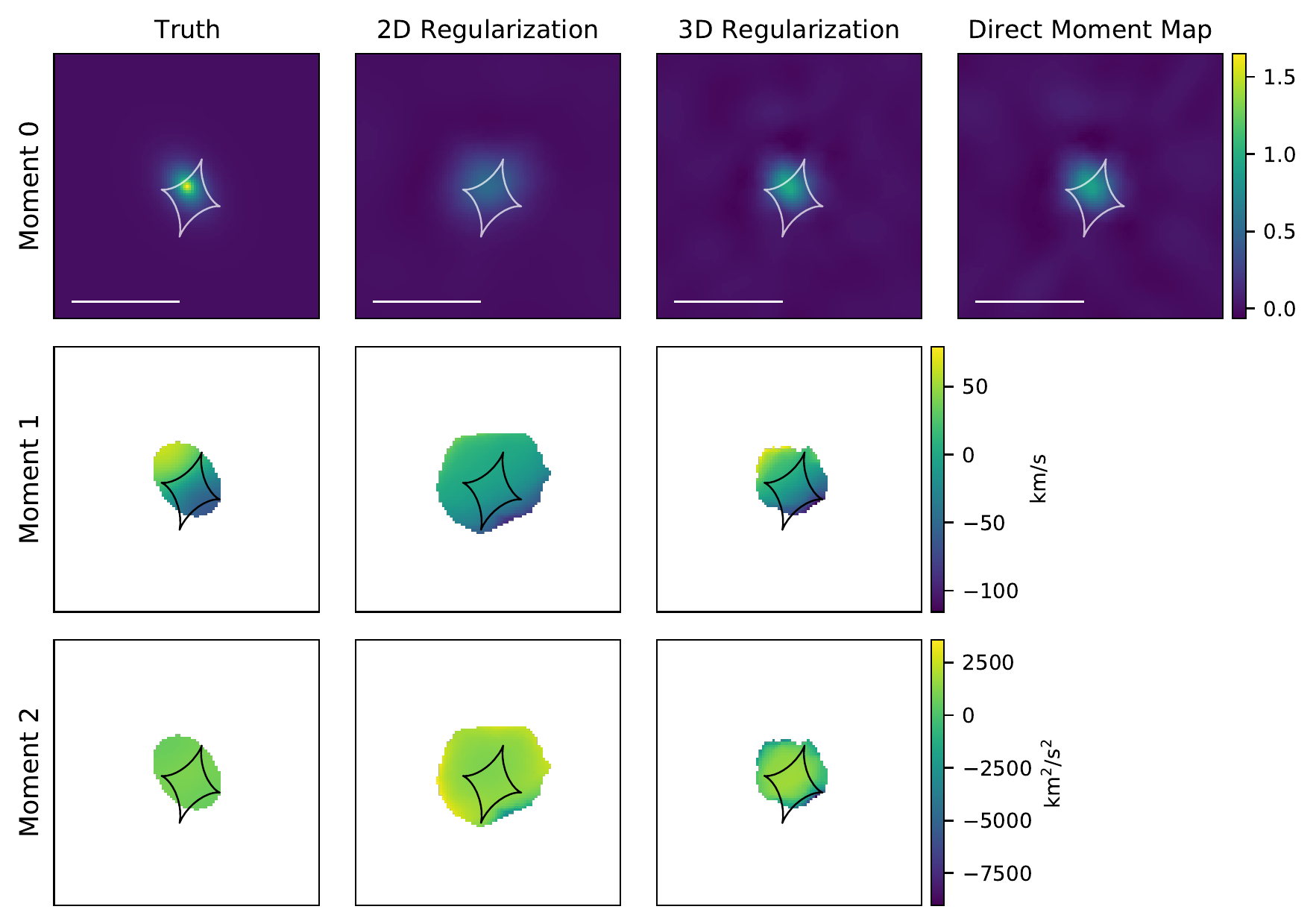}
		\caption{A comparison of the zeroth-, first-, and second-moment maps for the 
		various reconstruction techniques. Columns, left to right: the ``true" (input) source, the 2D-regularized 
		reconstruction, the 3D-regularized reconstruction, and the reconstruction derived directly from the observed 
		(lensed) moment map (see the text). Rows, top to bottom: the zeroth-, first-, and second-moment 
		maps of the reconstructions. The first- and second-moment maps are shown with masks 
		corresponding to $\textrm{S}/\textrm{N} \ge 3$ in the 2D and 3D zeroth-moment maps, while the truth 
		mask shows pixels that are $\ge 1/2$ the peak surface brightness. The ``direct" moment 
		map does not have corresponding higher-order moments because it does not reconstruct the full source cube. The 3D 
		zeroth-moment map reconstruction is similar in quality and effective resolution to the ``direct" 
		moment map reconstruction, while simultaneously recovering the full velocity structure of the cube. The greater 
		compactness of the channel-by-channel 3D reconstruction allows it to better reproduce the velocity gradient of 
		the true source compared to the 2D reconstruction.}\label{fig:full_moment_map_comparison}
	\end{figure*}

    We can also compare the various moment maps of the reconstructions to understand better how the full source 
	intensity and velocity fields are reconstructed. Figure \ref{fig:full_moment_map_comparison} shows the true source 
	alongside the zeroth-moment maps derived from the 2D- and 3D-regularized reconstructions shown in Figure 
	\ref{fig:source_truth_comparison}, as well as the reconstruction obtained from modeling the zeroth-moment map of 
	the data directly. (We will refer to the latter as the ``direct" zeroth-moment map reconstruction. Note that this 
	reconstruction lacks all spectral information that was present in the data cube.) The 2D-regularized and ``direct" 
	moment maps differ because individual channels of the data cube have lower S/N than the zeroth-moment map. The 
	lower S/N in the channels results in higher (per-channel) values of $\lambda$, and thus a lower effective 
	resolution, compared to the single value of $\lambda$ used for the ``direct" moment map reconstruction. Because of 
	the different form of the regularization, the 3D source achieves an effective spatial resolution similar to that of 
	the ``direct" moment map reconstruction by fitting to the full cube simultaneously, while also reconstructing the 
	entire velocity field of the source.
    
    Comparing the higher-order moment maps in rows 2 and 3 of Figure \ref{fig:full_moment_map_comparison}, we can see 
	that the 3D regularization reconstructs the first-moment map much better. This improvement occurs because 
	the 2D regularization reproduces all of the channels less accurately (if at all). Most importantly, the 
	first-moment calculation is sensitive to the emission in the outer channels where the 2D regularization 
	performs the worst. We do not draw any conclusions from the comparison of the second-moment maps, due to 
	the general difficulty of interpreting such higher moments in the presence of strong velocity gradients.

	\subsubsection{Spectral Resolution}\label{sec:spectral_resolution}
	
	\begin{figure*}
		\centering
		\includegraphics[width=5in]{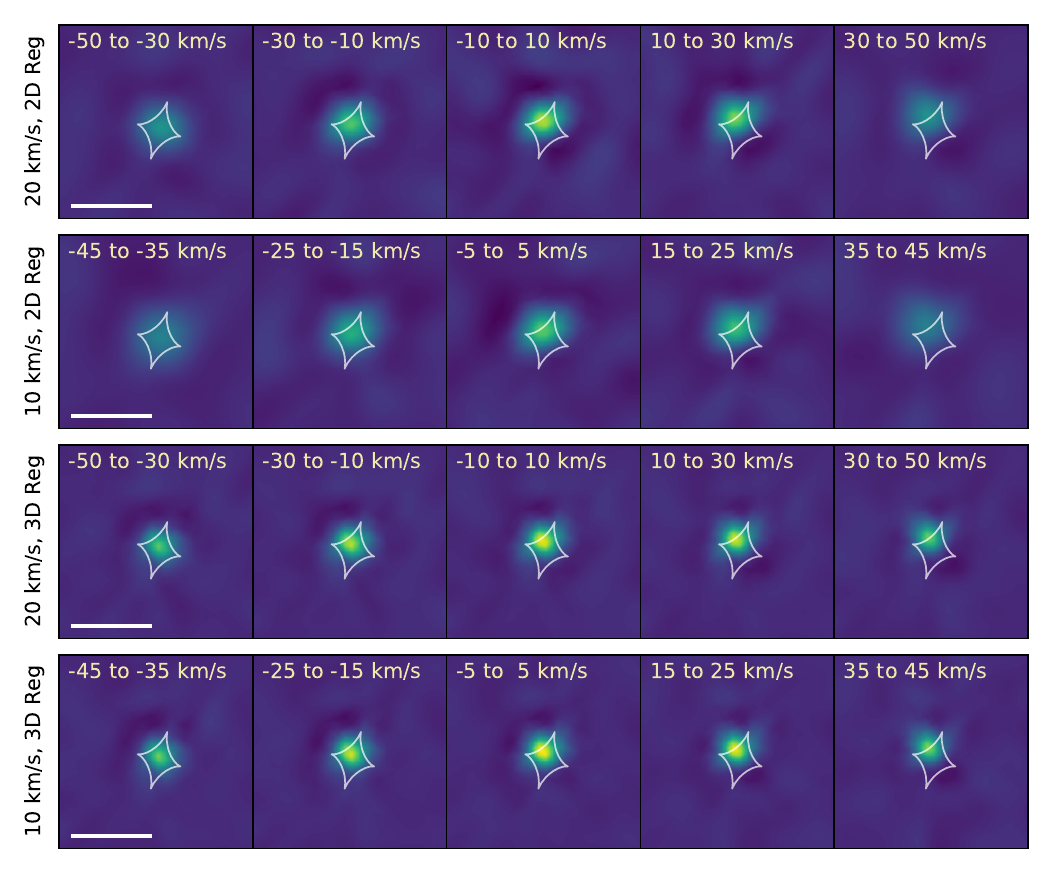}
		\caption{Select channel images using both reconstruction methods at two different spectral resolutions. The 
		top two rows show $20$ and $10 \,\kms$ resolutions, respectively, using 2D regularization, while the bottom two 
		rows show the same for 3D regularization. Each panel gives the velocity range of the channel shown. Note that 
		the two spectral resolutions have slightly different velocity ranges because of the alignment of the edges of 
		the channels, but the channels are chosen and re-scaled 
		to roughly match. Only a few channels near the line center are shown for clarity. Pixel values are in arbitrary 
		integrated flux units. The top two rows show how the compactness of the 2D reconstructions is dependent on the 
		channel width, while the bottom two rows show that the compactness of the 3D reconstructions is nearly 
		independent of channel width.} 
		\label{fig:spec_res_comparison}
	\end{figure*}
	
	To explore how the performance of the 3D regularization varies with spectral resolution, we simulate two versions of 
	the fiducial set of observations with the same sensitivity but two different spectral resolutions of $10$ and 
	$20 \,\kms$. Figure \ref{fig:spec_res_comparison} shows representative channel images of the difference between the 
	truth and the reconstructions for the two mock cubes and for both regularization methods. The 2D reconstructions are 
	sensitive to the choice of spectral resolution because narrower velocity channels result in channel images with lower 
	S/Ns. Because the 3D regularization fits all channels together, it maintains a consistent source size and 
	brightness profile regardless of the chosen channel width. This behavior is in line with the results of \citet{Chirivi2020}, who 
	find that their source parameter estimates are not negatively impacted (indeed, actually improve) as the number of channels 
	increases, even as the per-channel S/N decreases.
	
	\subsubsection{Spatial Resolution}\label{sec:spatial_resolution}
	\begin{figure}
		\centering
		\includegraphics[width=3.35in]{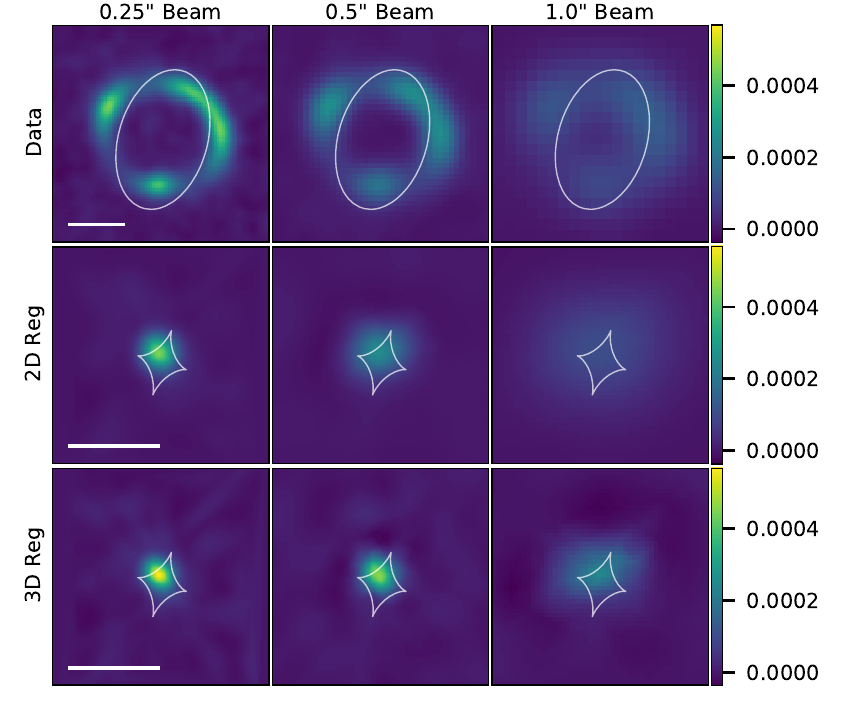}
		\caption{2D- and 3D-regularized reconstructions for three different spatial resolutions. Rows, top to bottom: 
		the zeroth-moment map of the mock data, the 2D-regularized reconstruction, and the 3D-regularized 
		reconstruction. Columns, left to right: mock data with $0.25''$, $0.5''$, and $1''$ beams, respectively.
		The 3D reconstruction is consistently more compact at all spatial resolutions. The white bar shows $1''$.}
		\label{fig:spatial_resolution_comparison}
	\end{figure}
	
	We illustrate the trade-off between spatial and spectral resolutions in Figure 
	\ref{fig:spatial_resolution_comparison}, where we show the moment maps of reconstructions derived from three 
	different sets of mock observations of the same source with $10 \,\kms$ velocity bins and $0.25$, $0.5$, and $1''$ 
	beams. Each of the data cubes has a peak S/N of 30. Both methods improve at higher spatial resolutions, 
	but the 3D reconstruction is consistently more compact with a higher peak surface brightness, even achieving a size 
	for a $0.5''$ beam comparable to that of the 2D source for a $0.25''$ beam. However, comparisons between spatial 
	resolutions are complicated by the fact that the same S/N does not correspond to the same real-world exposure 
	time at different resolutions.
	
	The above comparison suggests that it may be more valuable in many cases to prioritize spectral resolution over 
	spatial resolution to make the most effective use of available telescope time. Prospective users of 
	multi-configuration interferometers like ALMA, for example, must often make difficult decisions about how to achieve 
	compact synthesized beams $\sim \lambda/B_{\rm max}$: typical weather conditions limit the effective time available 
	for observations at short wavelengths (i.e., small $\lambda$), especially when the array is in an extended 
	configuration (i.e., the maximum projected baseline $B_{\rm max}$ is large). For a lensed source with a large
	velocity gradient, it may be easier to observe a longer-wavelength line in a more compact array configuration, as 
	long as the S/N per channel does not become prohibitively low.

	\section{Conclusions}\label{sec:conclusions}
	We have introduced modifications to a widely used Bayesian framework for reconstructing gravitationally lensed 
	sources, to facilitate the modeling of sources observed with integral field spectroscopy. Because of the nature of integral 
	field data, traditional (2D) regularization, which penalizes the squared gradient or Laplacian of each channel image 
	independently, results in reconstructions that have lower effective resolutions and are unable to reproduce the full 
	source velocity field accurately. Our newly introduced (3D) regularization scheme models the entire data cube 
	simultaneously, instead penalizing the squared gradient or Laplacian of the entire 3D surface brightness 
	distribution in order to account for the source velocity structure more effectively.
	
	We have investigated the performance of the new method and the properties of the derived sources using simulated 
	integral field observations of lensed rotating disks created using GalPaK$^{\text{3D}}$ and \code{lensmodel}. 
	We find that the 3D regularization leads to sources with higher effective resolutions and significantly better 
	reconstructions in the faint, high-velocity outer channels as compared to the 2D approach. The resulting 
	reconstructed 3D source cubes better reproduce both the zeroth- and first-moment maps of the 
	data, in addition to the individual channel images.
	
	The new form of the 3D regularization also changes the resolution and noise behavior in the source plane. Both the 
	(spatially varying) noise and resolution structures of the resulting source are nearly constant across channel 
	images in the 3D reconstruction compared to the spatial resolution and noise structure of the 2D reconstructions, 
	which vary strongly across channels. 
	
	We find that --- unlike the 2D reconstruction --- the 3D reconstruction is 
	robust to changes in the spectral resolution of the data. Along with the 
	better accuracy of the reconstructed sources, this fact incentivizes 
	obtaining higher spectral resolution observations of lensed sources even 
	at the same cost to the per-channel S/N ratio of the data. Applying 
	this new method to existing and future datasets will enable the extraction 
	of more information about the kinematic properties of lensed sources and 
	a better understanding of the underlying physics driving galaxies at high 
	redshift.

	\acknowledgments
	The authors thank Amitpal Tagore for his insights and generous assistance 
	with the \code{pixsrc} package, an anonymous referee for helpful comments 
	that improved the paper, and the National Science Foundation for support
	via grant AST-1716585.
	
	\software{
	\code{lensmodel} \citep{Keeton2011}, 
	\code{pixsrc} \citep{Tagore2014,Tagore2016}, 
	GalPaK$^{\text{3D}}$ \citep{Bouche2015}, 
	Numpy \citep{harris2020array}, 
	Scipy \citep{2020SciPy-NMeth}, 
	Astropy \citep{astropy:2013,astropy:2018}, 
	Matplotlib \citep{Hunter:2007}, 
	CHOLMOD \citep{CHOLMOD}
	}

	\bibliography{paper}

\begin{thebibliography}{}
\expandafter\ifx\csname natexlab\endcsname\relax\def\natexlab#1{#1}\fi

\bibitem[{{Astropy Collaboration} {et~al.}(2013){Astropy Collaboration},
  {Robitaille}, {Tollerud}, {Greenfield}, {Droettboom}, {Bray}, {Aldcroft},
  {Davis}, {Ginsburg}, {Price-Whelan}, {Kerzendorf}, {Conley}, {Crighton},
  {Barbary}, {Muna}, {Ferguson}, {Grollier}, {Parikh}, {Nair}, {Unther},
  {Deil}, {Woillez}, {Conseil}, {Kramer}, {Turner}, {Singer}, {Fox}, {Weaver},
  {Zabalza}, {Edwards}, {Azalee Bostroem}, {Burke}, {Casey}, {Crawford},
  {Dencheva}, {Ely}, {Jenness}, {Labrie}, {Lim}, {Pierfederici}, {Pontzen},
  {Ptak}, {Refsdal}, {Servillat}, \& {Streicher}}]{astropy:2013}
{Astropy Collaboration}, {Robitaille}, T.~P., {Tollerud}, E.~J., {et~al.} 2013,
  \aap, 558, A33

\bibitem[{{Astropy Collaboration} {et~al.}(2018){Astropy Collaboration},
  {Price-Whelan}, {Sip{\H{o}}cz}, {G{\"u}nther}, {Lim}, {Crawford}, {Conseil},
  {Shupe}, {Craig}, {Dencheva}, {Ginsburg}, {Vand erPlas}, {Bradley},
  {P{\'e}rez-Su{\'a}rez}, {de Val-Borro}, {Aldcroft}, {Cruz}, {Robitaille},
  {Tollerud}, {Ardelean}, {Babej}, {Bach}, {Bachetti}, {Bakanov}, {Bamford},
  {Barentsen}, {Barmby}, {Baumbach}, {Berry}, {Biscani}, {Boquien}, {Bostroem},
  {Bouma}, {Brammer}, {Bray}, {Breytenbach}, {Buddelmeijer}, {Burke},
  {Calderone}, {Cano Rodr{\'\i}guez}, {Cara}, {Cardoso}, {Cheedella}, {Copin},
  {Corrales}, {Crichton}, {D'Avella}, {Deil}, {Depagne}, {Dietrich}, {Donath},
  {Droettboom}, {Earl}, {Erben}, {Fabbro}, {Ferreira}, {Finethy}, {Fox},
  {Garrison}, {Gibbons}, {Goldstein}, {Gommers}, {Greco}, {Greenfield},
  {Groener}, {Grollier}, {Hagen}, {Hirst}, {Homeier}, {Horton}, {Hosseinzadeh},
  {Hu}, {Hunkeler}, {Ivezi{\'c}}, {Jain}, {Jenness}, {Kanarek}, {Kendrew},
  {Kern}, {Kerzendorf}, {Khvalko}, {King}, {Kirkby}, {Kulkarni}, {Kumar},
  {Lee}, {Lenz}, {Littlefair}, {Ma}, {Macleod}, {Mastropietro}, {McCully},
  {Montagnac}, {Morris}, {Mueller}, {Mumford}, {Muna}, {Murphy}, {Nelson},
  {Nguyen}, {Ninan}, {N{\"o}the}, {Ogaz}, {Oh}, {Parejko}, {Parley}, {Pascual},
  {Patil}, {Patil}, {Plunkett}, {Prochaska}, {Rastogi}, {Reddy Janga},
  {Sabater}, {Sakurikar}, {Seifert}, {Sherbert}, {Sherwood-Taylor}, {Shih},
  {Sick}, {Silbiger}, {Singanamalla}, {Singer}, {Sladen}, {Sooley},
  {Sornarajah}, {Streicher}, {Teuben}, {Thomas}, {Tremblay}, {Turner},
  {Terr{\'o}n}, {van Kerkwijk}, {de la Vega}, {Watkins}, {Weaver}, {Whitmore},
  {Woillez}, {Zabalza}, \& {Astropy Contributors}}]{astropy:2018}
{Astropy Collaboration}, {Price-Whelan}, A.~M., {Sip{\H{o}}cz}, B.~M., {et~al.}
  2018, \aj, 156, 123

\bibitem[{Birrer {et~al.}(2015)Birrer, Amara, \& Refregier}]{Birrer2015}
Birrer, S., Amara, A., \& Refregier, A. 2015, The Astrophysical Journal, 813,
  102

\bibitem[{Bouch{\'{e}} {et~al.}(2015)Bouch{\'{e}}, Carfantan, Schroetter,
  Michel-Dansac, \& Contini}]{Bouche2015}
Bouch{\'{e}}, N., Carfantan, H., Schroetter, I., Michel-Dansac, L., \& Contini,
  T. 2015, The Astronomical Journal, 150, 92

\bibitem[{Brewer \& Lewis(2006)}]{Brewer2006}
Brewer, B.~J., \& Lewis, G.~F. 2006, The Astrophysical Journal, 637, 608

\bibitem[{Chen {et~al.}(2008)Chen, Davis, Hager, \& Rajamanickam}]{CHOLMOD}
Chen, Y., Davis, T.~A., Hager, W.~W., \& Rajamanickam, S. 2008, ACM Trans.
  Math. Softw., 35, doi:10.1145/1391989.1391995

\bibitem[{{Chiriv{\`\i}} {et~al.}(2020){Chiriv{\`\i}}, {Y{\i}ld{\i}r{\i}m},
  {Suyu}, \& {Halkola}}]{Chirivi2020}
{Chiriv{\`\i}}, G., {Y{\i}ld{\i}r{\i}m}, A., {Suyu}, S.~H., \& {Halkola}, A.
  2020, \aap, 643, A135

\bibitem[{{Cresci} {et~al.}(2010){Cresci}, {Mannucci}, {Maiolino}, {Marconi},
  {Gnerucci}, \& {Magrini}}]{Cresci2010}
{Cresci}, G., {Mannucci}, F., {Maiolino}, R., {et~al.} 2010, \nat, 467, 811

\bibitem[{Dye \& Warren(2005)}]{Dye2005}
Dye, S., \& Warren, S.~J. 2005, The Astrophysical Journal, 623, 31

\bibitem[{Geach {et~al.}(2018)Geach, Ivison, Dye, \& Oteo}]{Geach2018}
Geach, J.~E., Ivison, R.~J., Dye, S., \& Oteo, I. 2018, The Astrophysical
  Journal Letters, 866, L12

\bibitem[{{Genzel} {et~al.}(2017){Genzel}, {F{\"o}rster Schreiber},
  {{\"U}bler}, {Lang}, {Naab}, {Bender}, {Tacconi}, {Wisnioski}, {Wuyts},
  {Alexander}, {Beifiori}, {Belli}, {Brammer}, {Burkert}, {Carollo}, {Chan},
  {Davies}, {Fossati}, {Galametz}, {Genel}, {Gerhard}, {Lutz}, {Mendel},
  {Momcheva}, {Nelson}, {Renzini}, {Saglia}, {Sternberg}, {Tacchella},
  {Tadaki}, \& {Wilman}}]{Genzel2017}
{Genzel}, R., {F{\"o}rster Schreiber}, N.~M., {{\"U}bler}, H., {et~al.} 2017,
  \nat, 543, 397

\bibitem[{{Genzel} {et~al.}(2020){Genzel}, {Price}, {{\"U}bler}, {F{\"o}rster
  Schreiber}, {Shimizu}, {Tacconi}, {Bender}, {Burkert}, {Contursi}, {Coogan},
  {Davies}, {Davies}, {Dekel}, {Herrera-Camus}, {Lee}, {Lutz}, {Naab}, {Neri},
  {Nestor}, {Renzini}, {Saglia}, {Schuster}, {Sternberg}, {Wisnioski}, \&
  {Wuyts}}]{Genzel2020}
{Genzel}, R., {Price}, S.~H., {{\"U}bler}, H., {et~al.} 2020, \apj, 902, 98

\bibitem[{Harris {et~al.}(2020)Harris, Millman, van~der Walt, Gommers,
  Virtanen, Cournapeau, Wieser, Taylor, Berg, Smith, Kern, Picus, Hoyer, van
  Kerkwijk, Brett, Haldane, del R{\'{i}}o, Wiebe, Peterson,
  G{\'{e}}rard-Marchant, Sheppard, Reddy, Weckesser, Abbasi, Gohlke, \&
  Oliphant}]{harris2020array}
Harris, C.~R., Millman, K.~J., van~der Walt, S.~J., {et~al.} 2020, Nature, 585,
  357

\bibitem[{{Hezaveh} {et~al.}(2017){Hezaveh}, {Perreault Levasseur}, \&
  {Marshall}}]{Hezaveh2017}
{Hezaveh}, Y.~D., {Perreault Levasseur}, L., \& {Marshall}, P.~J. 2017, \nat,
  548, 555

\bibitem[{Hunter(2007)}]{Hunter:2007}
Hunter, J.~D. 2007, Computing in Science \& Engineering, 9, 90

\bibitem[{{Jones} {et~al.}(2010){Jones}, {Swinbank}, {Ellis}, {Richard}, \&
  {Stark}}]{Jones2010}
{Jones}, T.~A., {Swinbank}, A.~M., {Ellis}, R.~S., {Richard}, J., \& {Stark},
  D.~P. 2010, \mnras, 404, 1247

\bibitem[{{Keeton}(2011)}]{Keeton2011}
{Keeton}, C.~R. 2011, {GRAVLENS: Computational Methods for Gravitational
  Lensing}, ascl:1102.003

\bibitem[{Kochanek \& Narayan(1992)}]{Kochanek1992}
Kochanek, C., \& Narayan, R. 1992, The Astrophysical Journal, 401, 461

\bibitem[{{Morningstar} {et~al.}(2019){Morningstar}, {Perreault Levasseur},
  {Hezaveh}, {Blandford}, {Marshall}, {Putzky}, {Rueter}, {Wechsler}, \&
  {Welling}}]{Morningstar2019}
{Morningstar}, W.~R., {Perreault Levasseur}, L., {Hezaveh}, Y.~D., {et~al.}
  2019, \apj, 883, 14

\bibitem[{Patr{\'{i}}cio {et~al.}(2018)Patr{\'{i}}cio, Richard, Carton,
  Contini, Epinat, Brinchmann, Schmidt, Krajnovi{\'{c}}, Bouch{\'{e}},
  Weilbacher, Pell{\'{o}}, Caruana, Maseda, Finley, Bauer, Martinez, Mahler,
  Lagattuta, Cl{\'{e}}ment, Soucail, \& Wisotzki}]{Patricio2018}
Patr{\'{i}}cio, V., Richard, J., Carton, D., {et~al.} 2018, Monthly Notices of
  the Royal Astronomical Society, 477, 18

\bibitem[{{Perreault Levasseur} {et~al.}(2017){Perreault Levasseur}, {Hezaveh},
  \& {Wechsler}}]{PerreaultLevasseur2017}
{Perreault Levasseur}, L., {Hezaveh}, Y.~D., \& {Wechsler}, R.~H. 2017, \apjl,
  850, L7

\bibitem[{{Refregier}(2003)}]{Refregier2003}
{Refregier}, A. 2003, \mnras, 338, 35

\bibitem[{{Riechers} {et~al.}(2008){Riechers}, {Walter}, {Brewer}, {Carilli},
  {Lewis}, {Bertoldi}, \& {Cox}}]{Riechers2008}
{Riechers}, D.~A., {Walter}, F., {Brewer}, B.~J., {et~al.} 2008, \apj, 686, 851

\bibitem[{Rizzo {et~al.}(2018)Rizzo, Vegetti, Fraternali, \& {Di
  Teodoro}}]{Rizzo2018}
Rizzo, F., Vegetti, S., Fraternali, F., \& {Di Teodoro}, E. 2018, Monthly
  Notices of the Royal Astronomical Society, 481, 5606

\bibitem[{{Rizzo} {et~al.}(2021){Rizzo}, {Vegetti}, {Fraternali}, {Stacey}, \&
  {Powell}}]{Rizzo2021}
{Rizzo}, F., {Vegetti}, S., {Fraternali}, F., {Stacey}, H.~R., \& {Powell}, D.
  2021, \mnras, 507, 3952

\bibitem[{{Scoville} {et~al.}(1997){Scoville}, {Yun}, \&
  {Bryant}}]{Scoville1997}
{Scoville}, N.~Z., {Yun}, M.~S., \& {Bryant}, P.~M. 1997, \apj, 484, 702

\bibitem[{Sharon {et~al.}(2019)Sharon, Tagore, Baker, Rivera, Keeton, Lutz,
  Genzel, Wilner, Hicks, Allam, \& Tucker}]{Sharon2019}
Sharon, C.~E., Tagore, A.~S., Baker, A.~J., {et~al.} 2019, The Astrophysical
  Journal, 879, 52

\bibitem[{Spilker {et~al.}(2018)Spilker, Aravena, B{\'{e}}thermin, Chapman,
  Chen, Cunningham, {De Breuck}, Dong, Gonzalez, Hayward, Hezaveh, Litke, Ma,
  Malkan, Marrone, Miller, Morningstar, Narayanan, Phadke, Sreevani, Stark,
  Vieira, \& Wei{\ss}}]{Spilker2018}
Spilker, J.~S., Aravena, M., B{\'{e}}thermin, M., {et~al.} 2018, Science, 361,
  1016

\bibitem[{Stark {et~al.}(2008)Stark, Swinbank, Ellis, Dye, Smail, \&
  Richard}]{Stark2008}
Stark, D.~P., Swinbank, A.~M., Ellis, R.~S., {et~al.} 2008, Nature, 455, 775

\bibitem[{Suyu {et~al.}(2006)Suyu, Marshall, Hobson, \& Blandford}]{Suyu2006}
Suyu, S.~H., Marshall, P.~J., Hobson, M.~P., \& Blandford, R.~D. 2006, Monthly
  Notices of the Royal Astronomical Society, 371, 983

\bibitem[{Tagore \& Jackson(2016)}]{Tagore2016}
Tagore, A.~S., \& Jackson, N. 2016, Monthly Notices of the Royal Astronomical
  Society, 457, 3066

\bibitem[{Tagore \& Keeton(2014)}]{Tagore2014}
Tagore, A.~S., \& Keeton, C.~R. 2014, Monthly Notices of the Royal Astronomical
  Society, 445, 694

\bibitem[{{Tikhonov} {et~al.}(1995){Tikhonov}, {Goncharsky}, {Stepanov}, \&
  {Yagola}}]{Tikhonov1995}
{Tikhonov}, A.~N., {Goncharsky}, A., {Stepanov}, V.~V., \& {Yagola}, A.~G.
  1995, Numerical Methods for the Solution of Ill-Posed Problems (Springer,
  Dordrecht)

\bibitem[{{Vegetti} \& {Koopmans}(2009)}]{Vegetti2009}
{Vegetti}, S., \& {Koopmans}, L.~V.~E. 2009, \mnras, 392, 945

\bibitem[{Virtanen {et~al.}(2020)Virtanen, Gommers, Oliphant, Haberland, Reddy,
  Cournapeau, Burovski, Peterson, Weckesser, Bright, {van der Walt}, Brett,
  Wilson, Millman, Mayorov, Nelson, Jones, Kern, Larson, Carey, Polat, Feng,
  Moore, {VanderPlas}, Laxalde, Perktold, Cimrman, Henriksen, Quintero, Harris,
  Archibald, Ribeiro, Pedregosa, {van Mulbregt}, \& {SciPy 1.0
  Contributors}}]{2020SciPy-NMeth}
Virtanen, P., Gommers, R., Oliphant, T.~E., {et~al.} 2020, Nature Methods, 17,
  261

\bibitem[{{Wallington} {et~al.}(1996){Wallington}, {Kochanek}, \&
  {Narayan}}]{Wallington1996}
{Wallington}, S., {Kochanek}, C.~S., \& {Narayan}, R. 1996, \apj, 465, 64

\bibitem[{Warren \& Dye(2003)}]{Warren2003}
Warren, S.~J., \& Dye, S. 2003, The Astrophysical Journal, 590, 673

\end{thebibliography}

	\appendix
    \section{Including a Line-spread Function}\label{app:lsf}

    For a single 2D image, the full lensing operator $\lens$ is related to the 
    blurring operator (the 2D PSF) $\blur_{\rm P}$ and the lensing-only operator 
    $\lens'$ via
    \begin{align}
        \lens = \blur_{\rm P} \lens'.
    \end{align}
    For data with no LSF, the 3D lensing matrix is given by
    \begin{align}
        \Lhat \equiv \Identity_{N_c} \otimes \lens.
    \end{align}

    When the data have an LSF that spreads emission in the spectral direction, 
    we can no longer write $\Lhat$ in block-diagonal form. Instead, we must 
    rewrite $\Lhat$ in the more general form
    \begin{align}
        \Lhat_{\rm LSF} = \blurhat_{\rm P,L} \Lhat',
    \end{align}
    where $\Lhat' = \Identity_{\rm N_{\rm c}} \otimes \lens'$ is the lensing-only matrix for the 3D cube, and $\blurhat_{\rm P,L}$ computes the combined blurring of the entire cube by both the PSF and LSF.

    $\blurhat_{\rm P,L}$ is a $N_s N_c \times N_s N_c$ block diagonal matrix. The blocks along the diagonal contain the 2D PSF. The off-diagonal blocks then show how a signal is spread into other channels; in general, blocks further from the diagonal will have lower amplitudes determined by the magnitude of the LSF for the given channel spacing.

	\section{Derivation of the $\Qhat$ Matrix}\label{app:Qhat}
	
	For a data cube with surface brightness distribution $\mathbf{S}(x,y,v_z)$, the gradient and Laplacian can be 
	written as
	\begin{align}
	\nabla \mathbf{S}(x,y,v_z) = \nabla_{x,y} \mathbf{S}(x,y,v_z) + \nabla_{v_z} \mathbf{S}(x,y,v_z),
	\end{align}
	and
	\begin{align}
	\nabla^2 \mathbf{S}(x,y,v_z) = \nabla^2_{x,y} \mathbf{S}(x,y,v_z) + \nabla^2_{v_z} \mathbf{S}(x,y,v_z),
	\end{align}
	respectively. The $\nabla_{x,y} \mathbf{S}$ and $\nabla^2_{x,y} \mathbf{S}$ terms are computed by the (spatial) 
	$\HtwoD$ matrix, while the extra $\nabla_{v_z} \mathbf{S}$ and $\nabla^2_{v_z} \mathbf{S}$ terms are computed by 
	the $\Qhat$ matrix introduced in Section \ref{sec:3D_regularization}. \citet{Suyu2006} and \citet{Tagore2014} give 
	the terms in $\HtwoD$ for a regular and irregular grid, respectively.
	
	Because the channel images (by assumption) have the same masks applied, the pixel grids for each channel are 
	identical. Therefore, the $\nabla_{v_z} \mathbf{S}$ and $\nabla^2_{v_z} \mathbf{S}$ terms for a given pixel only 
	depend on the corresponding pixels in adjacent channels. For a pixel $(i,j,k)$, where $i$ and $j$ denote the indices 
	in the spatial directions and $k$ denotes the index for the spectral direction, we can compute the gradient and 
	Laplacian using a central difference scheme. The gradient of the surface brightness is given by
	\begin{align}
	\nabla_{v_z} \mathbf{S} (i,j,k) \propto \left[\mathbf{S}(i,j,k+1) - \mathbf{S}(i,j,k-1)\right] \hat{r}_{k-1,k},
	\end{align}
	where $\hat{r}_{k-1,k}$ is the vector pointing from channel $k-1$ to channel $k$. The Laplacian is given by
	\begin{align}
	\nabla^2_{v_z} \mathbf{S} (i,j,k) \propto \left[\mathbf{S}(i,j,k+1) + \mathbf{S}(i,j,k-1) - 2\mathbf{S}(i,j,k)\right].
	\end{align}
	Note that the proportionality constant in each relation is absorbed into the second 
	regularization strength, $\eta$ (see Section \ref{sec:3D_regularization}). 
	
	We can adjust these terms for the edge channels by using a forward or backward difference instead of a central 
	difference. The forward difference for the first derivative is given by
	\begin{align}
	    \nabla_{v_z} \mathbf{S} (i,j,k) \propto \left[\mathbf{S}(i,j,k+1) - \mathbf{S}(i,j,k)\right],
	\end{align}
	and the backward difference is given by
	\begin{align}
	    \nabla_{v_z} \mathbf{S} (i,j,k) \propto \left[\mathbf{S}(i,j,k) - \mathbf{S}(i,j,k-1)\right].
	\end{align}
	The forward difference for the second derivative is given by
	\begin{align}
	    \nabla^2_{v_z} \mathbf{S} (i,j,k) \propto \left[\mathbf{S}(i,j,k+2) - 2\mathbf{S}(i,j,k+1) + \mathbf{S}(i,j,k)\right],
	\end{align}
	and the backward difference is given by
	\begin{align}
	    \nabla^2_{v_z} \mathbf{S} (i,j,k) \propto \left[\mathbf{S}(i,j,k) - 2\mathbf{S}(i,j,k-1) + \mathbf{S}(i,j,k-2)\right].
	\end{align}

	\section{Resolution Analysis}\label{app:resolution}
	\begin{figure}
		\centering
		\includegraphics[height=5.5in]{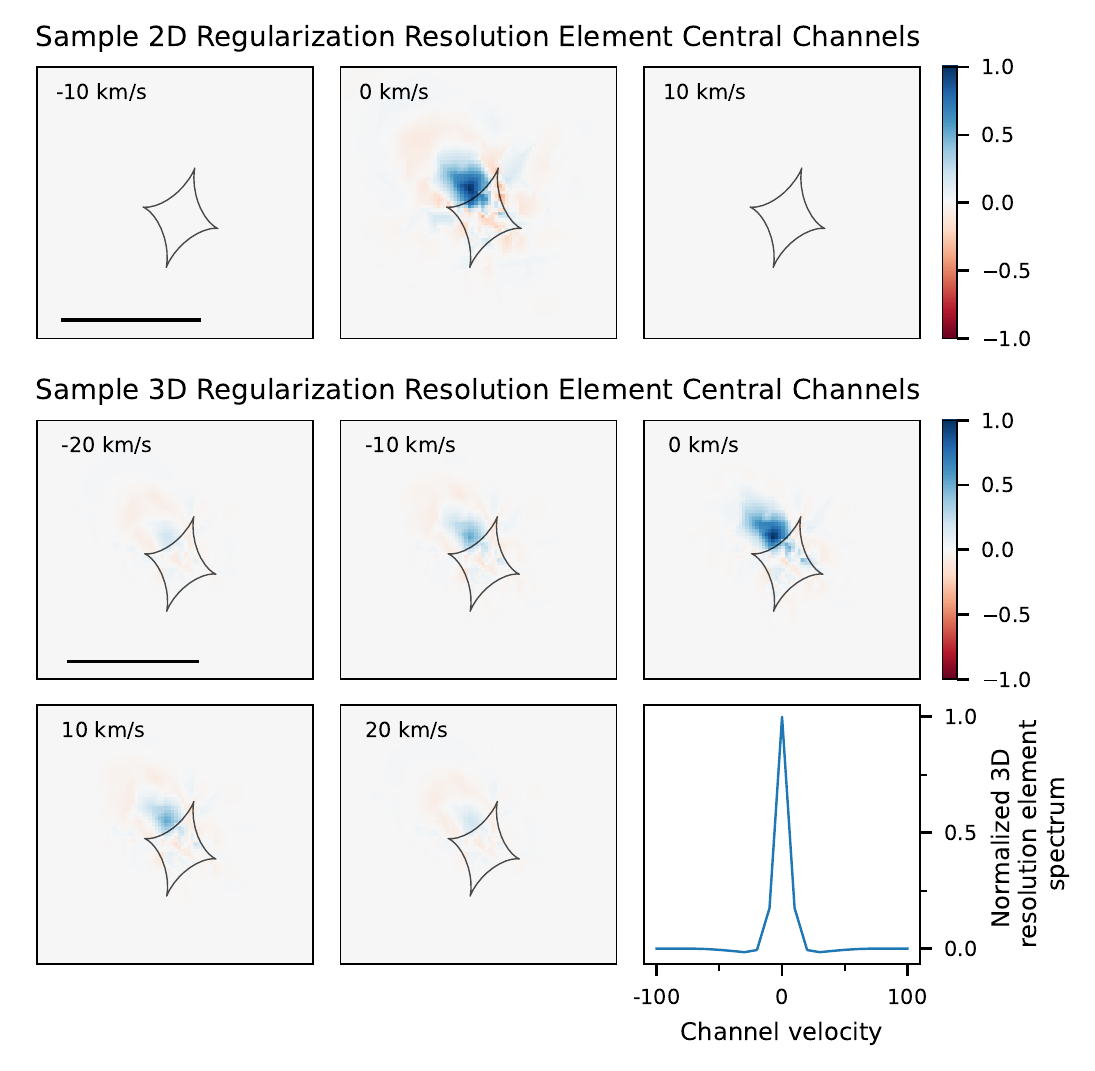}
	    \caption{
	    Example 3D effective resolution element in the source plane for
	    each reconstruction method. 
        Each beam corresponds to one column of the matrix
        $\Amat^{-1} \Fmat$ 
        in Equation \ref{eq:resolution_matrix}. The top row shows the 
        central three channels of the 2D effective resolution element. The 
        resolution element only appears in the central channel because the 
        reconstructed emission is not correlated between channels. The first 
        five panels in the bottom two rows show the five central channels of 
        the 3D effective resolution element, which contain most of the structure. The bottom right panel shows the spectrum  of the
        3D effective resolution element; here each channel is summed over
        the spatial pixels.
        Both resolution element cubes and the 3D beam spectrum are
        normalized to a peak of 1. 
        Compared to the 2D case, the 3D resolution element shows less
        spatial ``ringing,'' particularly in the central channel.
        However, the 3D regularization creates correlations in the
        spectral direction that give the reconstruction of a point source 
        an effective line width.}\label{fig:beam_example}
	\end{figure}
	
	A number of considerations make quantifying the source-plane resolution difficult in general and even more 
	difficult in the context of 3D regularization. First, regardless of the type of data, the variation of the magnification 
	factor across the source plane gives rise to a resolution element that changes with position. Second, the reconstruction 
	operation can also create ringing features (although regularization is designed to mitigate this) that result in 
	highly non-Gaussian beams. Finally, the resolution element for the 3D regularization is three-dimensional because of 
	the structure of the regularization matrix $\RthreeDhat$. Here we discuss some approaches to quantifying the 
	resolution in the source plane and provide an example that illustrates the general behavior of the different 
	resolution elements.

	One approach to modeling beams in the source plane is to create 
	mock observations with point sources, reconstruct those sources, 
	and fit Gaussians to measure beam properties \citep[e.g.,][]{Spilker2018}. As an alternative that does not require extensive reconstructions 
    with mock data, we can identify a matrix that describes the
    source-plane resolution elements. Consider starting with some
    source $\src_0$, creating a mock observation with noise, and then
    performing a reconstruction. The mock data can be written as
    $\data = \lens \src_0 + \noise$ where $\noise$ is a vector of
    noise in the data. Using Equations \ref{eq:s_ml} and \ref{eq:s_mp_2D},
    we can write the reconstructed (most probable) source as
    \begin{align}
        \src_{\text{MP}} &= \Amat^{-1} \lens^\top \cov^{-1} \data \nonumber \\
        &= \Amat^{-1} \lens^\top \cov^{-1} \lens \src_0 + \Amat^{-1} \lens^\top \cov^{-1} \noise \nonumber \\
        &= \Amat^{-1} \Fmat \src_0 + \Amat^{-1} \lens^\top \cov^{-1} \noise.
    \end{align}
    The reconstructed source is affected by the noise in the data. 
    Averaging over many realizations of the noise, 
    \begin{align}
        \langle \src_{\text{MP}} \rangle &= \langle\Amat^{-1} \Fmat \src_0\rangle + \langle\Amat^{-1} \lens^\top \cov^{-1} \noise\rangle \nonumber \\
        &= \Amat^{-1} \Fmat \src_0.
        \label{eq:resolution_matrix}
    \end{align}
    The first term is independent of noise, and the last term vanishes
    because $\langle\Mmat\noise\rangle=0$ 
    for any linear operator $\Mmat$. The matrix
    $\Amat^{-1} \Fmat$ thus 
    represents the combined actions of 
    lensing and computing the noise-averaged reconstruction $\langle \src_{\text{MP}} \rangle$ for a given source $\src_0$. If we let $\src_0$ represent individual source
    pixels, we can see that the columns of $\Amat^{-1} \Fmat$
    give the noise-averaged reconstructions of a set of point sources
    located at the vertices of the source-plane grid. These column vectors
    are the effective resolution elements in the source plane. This
    argument holds for the 3D case as well 
    if the matrices are replaced by
    their hatted counterparts.

    For 2D regularization $\Ahat^{-1} \Fhat$ is 
	block-diagonal, so each resolution element is confined to only one channel. However, 3D 
	regularization produces entries in the off-diagonal blocks of $\Ahat^{-1} \Fhat$ 
	that manifest as components of the resolution element in neighboring channels. The 
	differences between the 
	2D and 3D resolution elements can be seen in 
    Figure \ref{fig:beam_example}, which shows 
    examples appropriate for
    a position near the caustic 
	that is centered on the central channel of the cube. 
 
	The 2D regularization resolution element shown in Figure \ref{fig:beam_example} is for the
	central (highest S/N) channel of the source plane cube. The comparison of 
	this channel resolution element to the 3D regularization resolution element is the most favorable for 
	the 2D regularization, since the 2D resolution elements become significantly larger at 
	velocities farther from systemic, due to their lower S/N. The higher spatial resolution of the 3D regularization 
	across all of the channels
	appears to be the driver of the better source reconstruction shown in Figure 
	\ref{fig:source_truth_comparison}. The 3D regularization resolution element shows reduced spatial ringing in the central 
    channel, but also shows emission spread into neighboring channels. In this 
    case, most of the emission is confined to the central five channels 
    ($50 \,\text{km}\,\text{s}^{-1}$). The spectral width of the 3D 
    regularization resolution element depends on the spectral regularization 
    strength $\eta$.

\end{document}